\documentclass[sigconf]{acmart}

\AtBeginDocument{%
  }

\setcopyright{none}
\renewcommand\footnotetextcopyrightpermission[1]{}
\acmDOI{}
\acmISBN{}

\usepackage{listings}
\usepackage{xspace}
\usepackage{tikz}
\usepackage{amsmath}
\usepackage[loadonly]{enumitem}
\usepackage{booktabs}
\usepackage{multirow} 
\usepackage{longtable}
\usepackage{pifont}
\usepackage{tikz}
\usepackage{cleveref}
\usepackage{paralist}
\usepackage{algorithm2e}

\usepackage{comment}

\usepackage{tabularx}
\usepackage{xcolor}
\usepackage{listings}
\newcommand{\code}[1]{\lstinline[basicstyle=\ttfamily]{#1}}
\usepackage{array}
\usepackage{fontawesome5}
\usepackage{booktabs,array,fontawesome5}
\usepackage{colortbl}
\usepackage{adjustbox}

\RestyleAlgo{ruled} 
\LinesNumbered 

\newcommand{\todo}[1]{{\color{red} TODO: #1}}

\newcommand{\ayushi}[1]{{\color{orange} AS: #1}}
\newcommand{\fixed}[1]{{\color{brown} Fixed}}

\definecolor{stagepre}{HTML}{E6F4EA}
\definecolor{stageduring}{HTML}{E8F0FE}
\definecolor{stagepost}{HTML}{FFF4E5}
\definecolor{stagedeploy}{HTML}{FDE7E9}
\definecolor{integrated}{HTML}{228B22}   

\newcommand{\Paragraph}[1]{\vspace{.3em}\noindent\textbf{#1}}

\newcommand{\landseer}{\textsc{Landseer}\xspace}

\newcommand\sys{$\mathsf{Landseer}$\xspace}

\newcommand\defenses{TML defenses\xspace}
\newcommand\TML{TML\xspace}
\newcommand\mlops{MLOps\xspace}

\newcommand{\atdefense}{\textbf{T}\textsuperscript{ev}}
\newcommand{\ordefense}{\textbf{T}\textsuperscript{ou}}
\newcommand{\dpdefense}{\textbf{T}\textsuperscript{dp}}
\newcommand{\wmdefense}{\textbf{T}\textsuperscript{wm}}
\newcommand{\fpdefense}{\textbf{T}\textsuperscript{fp}}
\newcommand{\fairnessdefense}{\textbf{T}\textsuperscript{gf}}
\newcommand{\expdefense}{\textbf{T}\textsuperscript{ex}}

\newcommand{\stagepre}{\colorbox{stagepre}{Pre}}
\newcommand{\stageduring}{\colorbox{stageduring}{During}}
\newcommand{\stagepost}{\colorbox{stagepost}{Post}}
\newcommand{\stagedeploy}{\colorbox{stagedeploy}{Deploy}}

\newcommand{\tool}[3]{%
  \ensuremath{\textsc{#1}_{\text{\scriptsize\sffamily #2}}^{\text{\scriptsize #3}}}%
}
\newcommand{\AR}[2]{\tool{AR}{#1}{#2}}
\newcommand{\OR}[2]{\tool{OR}{#1}{#2}}
\newcommand{\WM}[2]{\tool{WM}{#1}{#2}}
\newcommand{\FP}[2]{\tool{FP}{#1}{#2}}
\newcommand{\DP}[2]{\tool{DP}{#1}{#2}}
\newcommand{\GF}[2]{\tool{GF}{#1}{#2}}
\newcommand{\EX}[2]{\tool{EX}{#1}{#2}}
\newcommand{\T}[2]{\tool{T}{#1}{#2}}

\usepackage[most]{tcolorbox}
\usepackage{xcolor}

\definecolor{findingbg}{RGB}{255,250,235}
\definecolor{findingborder}{RGB}{180,120,20}

\newtcolorbox{findingBox}[1][]{
  enhanced,
  breakable,
  colback=findingbg,
  colframe=findingborder,
  boxrule=0pt,
  frame hidden,
  borderline west={2pt}{0pt}{findingborder},
  sharp corners,
  left=6pt,
  right=6pt,
  top=4pt,
  bottom=4pt,
  before skip=6pt,
  after skip=6pt,
  #1
}
\newcommand{\numselectedtools}{35\xspace}
\newcommand{\nummldefences}{7\xspace}
\newcommand{\numcombinations}{700\xspace}
\newcommand{\numnoartifacts}{11\xspace}
\newcommand{\numwithartifacts}{24\xspace}
\newcommand{\numofficialartifacts}{22\xspace}
\newcommand{\numthirdparty}{2\xspace}

\newcommand{\numintegrated}{11\xspace}

\begin{document}

\title{Landseer: Exploring the Machine Learning Defense Landscape}

\author{Ayushi Sharma}
\authornote{Both authors contributed equally to this research.}
\affiliation{%
  \institution{Purdue University}
  \city{West Lafayette}
  \country{USA}
}
\email{sharm616@purdue.edu}

\author{Rosemary Agbozo}
\authornotemark[1]
\affiliation{%
  \institution{Purdue University}
  \city{West Lafayette}
  \country{USA}
}
\email{ragbozo@purdue.edu}

\author{Santiago Torres-Arias}
\affiliation{%
  \institution{Purdue University}
  \city{West Lafayette}
  \country{USA}
}
\email{santiagotorres@purdue.edu}

\author{Zahra Ghodsi}
\affiliation{%
  \institution{Purdue University}
  \city{West Lafayette}
  \country{USA}
}
\email{zahra@purdue.edu}
\renewcommand{\shortauthors}{Sharma et al.}

\begin{abstract}

Machine learning systems face diverse threats that undermine robustness, privacy, and fairness. Although many defenses have been proposed, each typically addresses a single risk in isolation. Real-world deployments, however, require these defenses to be composed to meet multiple guarantees simultaneously. The process of composing defenses is complex and not well understood, and its impact on performance and security remains unclear.

We present \landseer, a modular framework for integrating machine learning (ML) defenses into the ML lifecycle and systematically evaluating their composition. \landseer encapsulates defenses as containerized modules, allowing existing and new techniques to be plugged in with minimal effort. Its evaluation engine automates experiments across multiple metrics, supporting the study of defenses both individually and in combination. In a preliminary study, we identified \numselectedtools state-of-the-art machine learning defenses. After filtering for reproducibility, we analyzed their performance using Landseer’s unified evaluation process. 

Our findings reveal gaps in replicability across defense families and provide insights into the challenges and opportunities in integrating multiple defenses, establishing a foundation for improving the reliability of machine learning systems.

\end{abstract}


\maketitle
\pagestyle{plain}

\section{Introduction}

The increasing popularity of Machine Learning (ML) systems in critical applications and industries such as cybersecurity~\cite{apruzzese2023role, amer2020dynamic}, healthcare~\cite{bravi2024development, rahmani2021machine, khayyam2020artificial}, and finance~\cite{sirignano2021universal}
has raised growing concerns about the susceptibility of these systems to various types of attacks and trust concerns.
Notable incidents include physical evasion of traffic sign detectors using small stickers~\cite{eykholt2018robustphysicalworldattacksdeep}, adversarial perturbations that can alter medical image diagnoses ~\cite{tsai2023adversarial}, and data poisoning attacks that lead to backdoors~\cite{Zhang_2024_CVPR}. 
These threats target various trustworthy ML dimensions including performance~\cite{knauer2024phantom, wang2023analysis}, robustness~\cite{finlayson2019adversarial,newaz2020adversarial}, privacy~\cite{shokri2017membership,liu2022membership,choquette2021label,yeom2018privacy,fredrikson2015model,melis2019exploiting}, and fairness~\cite{lewicki2023out,rezaei2025fairness,ueda2024fairness}, among others.

In response, industry and academia have developed defense techniques~\cite{8406613, mironov2017renyi, cohen2024simple} to prevent attacks targeting any of the trustworthy machine learning qualities described above.
For example, robustness risks can be addressed by adversarial training~\cite{Zhang2019TheoreticallyPT, Zheng2019EfficientAT}, and backdoors can be mitigated via data sanitization ~\cite{Borgnia2020StrongDA}. 
Each defense is typically designed to be applied at a specific stage of the ML lifecycle and can serve as a building block for
constructing secure ML systems. 
In real-world scenarios, multiple risks or attack vectors are often present simultaneously. 
As a result, it may be necessary to incorporate various defense strategies to address a wide range of security and reliability requirements. 
Consequently, we need the ability to combine defenses while keeping their purpose and effectiveness intact.

Unfortunately, combining multiple ML defenses is not trivial. 
Composability of defenses is not well understood, and the aggregate effect on system performance and security guarantees is difficult to predict. 
This difficulty is due to the diversity of existing techniques. 
For any one risk class, a variety of defenses exist that are applied at different stages of development or deployment and are configured by a unique set of parameters.
As a result, the landscape of possible defenses that could be applied to an ML system grows with combinatorial complexity, making the impact of combining multiple trustworthy machine learning defenses challenging to evaluate.

Due to the aforementioned challenges, prior efforts to combine ML defenses and study their interactions and effects on system performance and risks have been mainly limited to pairwise evaluations consisting of two ML defenses at a time~\cite{gittens2022adversarial, chen2023privacy,noppel2024sok, szyller2023conflicting, duddu2024sok, tran2024effects}.
Duddu et al.~\cite{duddu2024sok} studied the interactions between defenses and risks, examining the unintended effect of one defense on other risks (not targeted by the defense). 
However, this work does not offer any insight into how defenses interact with each other.
In a follow-up, Duddu et al.~\cite{duddu2024combining} proposed a non-empirical model to predict if combining specific existing defenses will be effective. 
However, the model is primarily proposed for pairwise combinations, which limits expandability when multiple defenses are considered in the same or different stages of the pipeline.  

In this paper, we close this gap by developing \landseer, a comprehensive framework to integrate and systematically analyze ML defenses in a typical ML pipeline to discover automated insights on defense interactions. 
We define and explore the notion of \emph{composability} for ML defenses, investigating structural requirements for a defense to coexist with other tools in the pipeline, as well as qualitative interference between defenses that affect trust metrics. 
\landseer's architecture consists of an onboarding module, a pipeline explorer, and an interference analysis engine. 
The onboarding module takes a tool description and corresponding software artifact and generates a reproducibility record together with a structurally composable artifact. 
The pipeline explorer is the core empirical component that collects data on qualitative composability of defense combinations. 
The interference analysis engine performs root-cause analysis on collected data to identify sources of interference and provide additional insights about the tools and their interactions.

We demonstrate the capabilities of \landseer by surveying \numselectedtools defense techniques from literature and examining their composability. 
Based on this survey, we identify integration barriers and compile guidelines for researchers to avoid composability obstacles. 
From the examined tools, we onboard \numintegrated and demonstrate that \landseer is able to reproduce existing results on interactions of defenses, and can further surface new insights. 
Notably, our results show that defense interactions depend on several factors, including the stage at which a defense is applied and the order of defenses within a stage, which have not been examined in prior work. 
\landseer can help ML practitioners identify best-performing combinations according to a security policy, and enable ML researchers to extract new insights into ML defense interactions to build easy-to-integrate defenses and lower barriers for their adoption. 



Our contributions can be summarized as follows:

\begin{itemize}
    \item \textbf{Define and study the notion of composability} of various Trustworthy Machine Learning (TML) techniques in a Machine Learning Pipeline. 
    Our taxonomy outlines two composability criteria, structural and qualitative, that describe a defense's ability to be integrated in an ML deployment.

    \item \textbf{Design and implement \landseer}, a generalizable framework to adopt and analyze the composability of defenses.
    
    \item \textbf{Test a corpus of \numselectedtools \TML techniques} against \landseer to replicate previous literature on \TML defense interactions, as well as discover new adverse interactions (or interferences) among tools across \numcombinations total combinations.
    
    \item \textbf{Identify new insights to aid in tool composability} through two case studies to help practitioners and researchers in future tool development.

\end{itemize}

\section{Background}\label{sec:background}
\subsection{Machine Learning Operations} 

Machine learning operations (\mlops) describe the activities and lifecycle to create and deploy machine learning models.
Example operations include collection and processing of data, the use of this data to train a model, and deployment of the model to users. 
At a high level, we can categorize the \mlops lifecycle into the following stages: (i) \emph{data processing} involves collecting a dataset and processing it through cleaning and transformation operations, (ii) \emph{training} consists of designing and training an ML model over the collected training data, (iii) \emph{testing} carries out the evaluation of the resulting model from the previous stage on a holdout set of data across a set of metrics and might involve further adjustments to the model, and (iv) \emph{deployment} where the model is deployed at scale and performs computations over production data.

\subsection{Trustworthy Machine Learning Techniques}
\label{section:defense_types} 
ML systems are designed to perform well (for example, have high prediction accuracy) on never-before-seen data. 
However, ML systems should also exhibit other desirable characteristics for deployment in the real world, such as robustness, fairness, and privacy protection. 
We provide an overview of Trustworthy Machine Learning (\TML) techniques that we study in this work, and present relevant definitions and evaluation metrics
summarized in Table~\ref{tbl:evalmetric}.

\noindent\textbf{Adversarial Robustness.}
An adversarial evasion attack deceives the model into making incorrect predictions by adding carefully crafted perturbations to input~\cite{szegedy2013intriguing, goodfellow2014explaining}. 
Evasion robustness techniques $\textbf{T}^\textbf{ev}$ aim to suppress the effects of adversarial examples by improving model resilience.
Methods to achieve robustness include adversarial training
~\cite{Madry2017TowardsDL, Shafahi2019AdversarialTF}, randomized smoothing
~\cite{Cohen2019CertifiedAR}, and input data transformations
to strip perturbations~\cite{Guo2018CounteringAI, cohen2024simple, xu2017feature}.
Evaluating the robustness of a model can be performed by testing the model against a dataset containing adversarial examples. 
We use the adversarial robustness accuracy $\textbf{m}^\textbf{ev}$ as the evaluation metric measuring the percentage of adversarial examples correctly classified. 

\noindent\textbf{Outlier Robustness.}
Outliers are data points (accidentally or maliciously added) that deviate from the distribution of other points.
Techniques to improve outlier robustness $\textbf{T}^\textbf{ou}$ include
dataset sanitization~\cite{Borgnia2020StrongDA, Han2018CoteachingRT, steinhardt2017certified} and model-focused purification~\cite{Wu2021AdversarialNP, Li2023ReconstructiveNP, dolan2018fine}.
Depending on the technique, we use 
attack success rate $\textbf{m}^\textbf{ou}_\textbf{asr}$ 
~\cite{biggio2012poisoning} or the area under the receiver operating characteristic curve $\textbf{m}^\textbf{ou}_\textbf{ar}$ 
~\cite{davis2006relationship} as metrics to evaluate robustness to outliers.

\begin{table}[t]
\centering
\caption{Description of ML risks and corresponding \TML techniques and metrics for evaluating the effectiveness of each technique. $\uparrow$ ($\downarrow$) indicates a higher (lower) value is better.}
\renewcommand{\arraystretch}{.5} 
\begin{tabular}{ll p{2.8cm}}
\toprule
Risk & \TML Technique & Eval. Metric \\
\midrule
Evasion & Adv. Robustness $\atdefense$ & Adv. robustness acc. $\textbf{m}^\textbf{ev} (\uparrow)$\\
Poisoning & Outl. Robustness $\ordefense$& Att. success rate  $\textbf{m}^\textbf{ou} (\downarrow)$, area under ROC curve $\textbf{m}^\textbf{ar} (\uparrow)$\\
Privacy & Diff. privacy $\dpdefense$& DP parameters  $\textbf{m}^\textbf{dp}_\mathbf{\epsilon} (\downarrow)$ and $\textbf{m}^\textbf{dp}_\mathbf{mia} (\downarrow)$\\
Intel. Prop. & Watermarking $\wmdefense$& Watermark detection accuracy $\textbf{m}^\textbf{wm} (\uparrow)$\\
Intel. Prop. & Fingerprinting $\fpdefense$& Fingerprint detection confidence score $\textbf{m}^\textbf{fp} (\uparrow)$\\
Bias & Group Fairness $\fairnessdefense$& Demographic parity $\textbf{m}^\textbf{fa} (\uparrow)$\\
Transparency & Explanation $\expdefense$& Explanation error $\textbf{m}^\textbf{ex} (\uparrow)$\\
\bottomrule
\end{tabular}
\label{tbl:evalmetric}

\end{table}

\noindent\textbf{Privacy.}
Privacy attacks include membership~\cite{shokri2017membership}, attribute~\cite{mehnaz2020black}, and reconstruction~\cite{carlini2019secret} attacks, which aim to leak the usage of a data point or its hidden attributes or recreate them from the model's output.
Differential Privacy (DP) is a popular technique $\textbf{T}^\textbf{dp}$ to protect privacy by adding controlled yet random noise to the data while enabling the learning task simultaneously. DP mechanisms can be designed and evaluated based on 
the privacy budget $\textbf{m}^\textbf{dp}_\mathbf{\epsilon}$ (indicates the strength of the privacy guarantee)
, and  $\textbf{m}^\textbf{dp}_\mathbf{mia}$~\cite{Abadi2016DeepLW, mironov2017renyi} (quantifying susceptibility to membership inference attacks).

\noindent\textbf{Watermarking.}
Adversaries may attempt to steal ML models to avoid the high training or continuous per query costs~\cite{10.1145/3595292, tramer2016stealing}. 
Watermarking techniques $\textbf{T}^\textbf{wm}$ embed ownership information into trained models that can be later verified via trigger/backdoor marks using watermarked training data or after-training finetuning 
~\cite{Adi2018TurningYW}, training-time embeddings
~\cite{Uchida2017EmbeddingWI}, API-time dynamic watermarking 
~\cite{szyller2021dawn}, and data-centric tracing 
~\cite{sablayrolles2020radioactive}. 
%
%
The watermark accuracy metric $\textbf{m}^\textbf{wm}$ evaluates the presence of watermarks in a model and the percentage (or fraction) of the correctly identified watermarks.

\noindent\textbf{Fingerprinting.}
Fingerprinting is a technique $\textbf{T}^\textbf{fp}$ to verify the ownership of an ML model by
analyzing inherent characteristics such as the model's decision-boundary signature
~\cite{Peng2022FingerprintingDN}.
To compute the accuracy of a fingerprint, we can use a confidence score metric $\textbf{m}^\textbf{fp}$ accompanying a prediction that a particular model contains a specific fingerprint
~\cite{Cao2019IPGuardPT}. 


\noindent\textbf{Fairness.}
ML models are susceptible to biases that would make their decisions \emph{unfair}~\cite{angwin2022machine,barocas2023fairness}.
In this paper, we consider the notion of group fairness, which ensures that different groups are treated equally~\cite{pessach2022review}. Techniques for group fairness ($\textbf{T}^\textbf{fa}$) include
rebalancing training data~\cite{Calmon2017OptimizedPF} or modifying the learning objective \cite{agarwal2018reductionsapproachfairclassification}
To evaluate fairness, we adopt the demographic parity (statistical parity) metric $\textbf{m}^\textbf{fa}$ to measure whether different groups receive positive outcomes at equal rates. 

\noindent\textbf{Explanation}
Understanding model decisions can help promote trust and fairness~\cite{lipton2018mythos}. Explanation techniques $\textbf{T}^\textbf{ex}$ aim to clarify how predictions are made by methods including examining how changes during training data affect outputs, 
or incorporating explanation terms directly into the model’s objective function~\cite{ribeiro2016should,lundberg2017unified}. 
We adopt the explanation error metric $\textbf{m}^\textbf{ex}$ which quantifies how accurately the explanation reflects the model's true behavior ~\cite{petsiuk2018rise}. 


\section{Landseer Overview}
\label{sec:landseer_over}

Thus, \landseer is a \emph{framework} designed to explore the ways in which \defenses tools can be composed.
To do so, it relies on modules to perform comprehensive composability tests on available \defenses.
This section presents the definitions required to understand the framework, the flow users carry out to use it, and its three major building blocks.

\subsection{Definitions}
\label{subsec:compo_defin}

We begin by defining the following concepts:

\begin{itemize}
    \item \textbf{Structural Composability} specifies whether a tool can be integrated into a pipeline and produce an output, i.e, whether its inputs and outputs are compatible with adjacent components such that it can be seamlessly chained within a larger workflow.
    \item \textbf{Qualitative Composability} describes whether two or more defenses can coexist without significantly degrading overall system or defense performance. 
    
    \item \textbf{\defenses Interference} occurs when one tool changes the performance of another. When two tools are not qualitatively composable, they negatively interfere. To provide a deeper insight into how tools can fail to compose, \landseer also identifies \emph{interference patterns} (or types), which can ultimately assist practitioners in communicating the consequences of deploying \defenses together.

\end{itemize}

We will utilize these definitions throughout the rest of the paper to characterize how \defenses may (or may not) integrate. 

\subsection{\landseer Flow}

A user of \landseer, such as an ML engineer, is tasked with ensuring \defenses can be successfully integrated into their system.
To do so, they must first identify target \defenses tools denoted as a tuple of $< Code, Target>$.
$Code$ refers to an existing implementation of such a tool, whereas $Target$ is a vector containing a specific benchmark for \defenses metrics. We provide some examples for this in Section~\ref{sec:evaluation}.

Besides a collection of tools to test and integrate, a user must also provide a model $M$, a dataset $D$, and a hyperparameter set $H$.
In doing so, a user can verify whether a tool is composable in their particular setting.

\subsection{\landseer Components}
\landseer consists of three major components:

\begin{itemize}
    \item \textbf{Onboarding Module:} This module is tasked with converting \defenses into composable artifacts. 
    To do so, it requires a tool description (e.g, an academic paper) and its corresponding software artifact.
    From these, it builds a reproducibility record (i.e., their respective expected reproducibility values), runs a series of tests to confirm reproducibility and structural composability, and produces an artifact that can be integrated into an MLOps pipeline.
    
    \item \textbf{Pipeline Explorer:} After the tool(s) pass the onboarding module, the Pipeline Explorer module builds combinations to identify whether tools produce satisfactory results (i.e., are qualitatively composable, as described below).
    The pipeline explorer also verifies whether \defenses defenses work within a target dataset, model-architecture, and other user-defined parameters. 
    At a high level, this process can be an enumeration (or cross-product) of all available tools for a particular configuration, though \landseer applies various optimizations to avoid recomputing intermediate artifacts.

    \item \textbf{Interference Analysis:} Once all pipeline combinations are explored, the results are processed through an analysis engine, which provides feedback on which tool combinations are interfering, the type of interference, and additional insights about the tools.
\end{itemize}

\begin{figure*}[th!]
    \centering
    \includegraphics[trim=10 20 0 10, width=.95\textwidth]{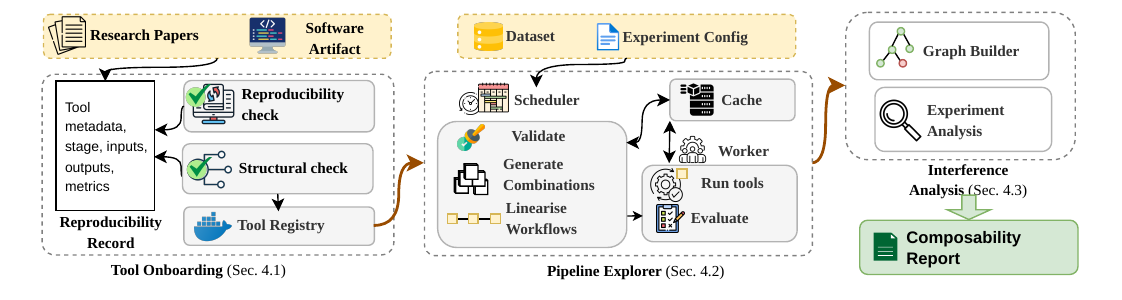}
    \caption{Overview of \landseer design. Practitioners define tools of interest in a domain-specific language, while machine learning researchers can integrate attacks and defenses to test against (Section~\ref{sec:litreview}). Results are produced as series of n-tuples describing a defense combination, a test accuracy, and a collection of \defenses scores (Section~\ref{sec:architecture}).}
    \label{fig:overview}
\end{figure*}

Figure~\ref{fig:overview} depicts these modules and the flow.
We will explain these components in detail in Section ~\ref{sec:architecture}.

\subsection{Threat Model}

\landseer is at its core a meta-framework for trustworthy machine learning composition.
As such, it inherits attacker goals and capabilities from the defenses integrated in our study.
Further, \landseer explicitly operates under a framework placing defense benchmarks against defense constructions, which directly encapsulate the relationship between attackers and defenders for a particular Trustworthy Machine Learning setting.
Lastly, we assume there are no emergent attack classes derived from threat composition that are not addressed by composing their corresponding defenses.

The security of the framework itself falls within standard distributed system assumptions under a non-byzantine setting.
Operators of a \landseer deployment are in full control of all the nodes in the infrastructure, and full compromise of a particular node will result in faulty computation.
Attacks that result in compromised input tooling (e.g., supply chain attacks through research code) are addressed under regular risk assessment methodologies and are out of scope for this paper.

\section{\landseer Architecture}
\label{sec:architecture}

In this section, we further develop the specific modules that are part of the \landseer framework.




\subsection{Onboarding Module}
\label{onboarding_module}

Before a \TML tool can be used in a Landseer pipeline, it must first be converted from a standalone research artifact into a reliable pipeline component. 
This step is necessary because \TML tools are often released with different assumptions about datasets, model architectures, dependencies, input formats, output formats, and evaluation scripts. 
Moreover, a tool may run correctly in its original setting (e.g., as a standalone Python script), but still be difficult to compose with other tools in an MLOps pipeline \cite{Canonne2020TheDG} (Refer Table \ref{tbl:litSurveyFA}). 
This is important because structural composability forms the baseline for qualitative composability -- Landseer can only evaluate how tools interact if they can be integrated into a pipeline.

In order to successfully adapt a \TML tool into a pipeline, \landseer must ensure that the following invariants hold:
In particular, it should determine: (i) whether the tool can be executed, (ii) whether it reproduces the behavior reported by the original artifact (e.g., it matches the results reported in the original paper), (iii) where the tool belongs in the ML lifecycle, and (iv) what input and output artifacts are required to chain it with other tools.

\subsubsection{Reproduction Records}
In order to track progress in adapting \TML tools into the pipeline, \landseer keeps a \emph{Reproduction Record} for each candidate tool. 
Succinctly put, this record stores both the evidence that the tool works in isolation and the interface information needed to use it inside a larger MLOps pipeline. 
A Reproduction Record contains three classes of information: artifact metadata, reproducibility outcome, and structural properties.

\paragraph{Artifact metadata} This includes the paper or artifact source, the code URL, the defense category, the intended pipeline stage, the supported datasets, the framework requirements, and the reported target metrics. 
These fields describe what the tool claims to do and where it is expected to fit in the trustworthy ML lifecycle.
Populating this information is a structured process that can be carried out by a human or an automated agent.

\paragraph{Reproducibility Outcome} This includes whether the artifact executes successfully, 
and what target metric was reproduced. 
This part of the record establishes the standalone baseline for the tool. 
Without this baseline, it is difficult to determine later whether a change in performance is caused by the composition with another tool or by an already unstable artifact.

\paragraph{Structural Interface} This includes the expected input artifact, the produced output artifact, the required auxiliary files, the parameter schema, and any format constraints on datasets, model checkpoints, or framework versions. 
These fields determine the tool's structural composability: whether another stage of the pipeline can consume its outputs without any transformation (e.g., between data representations).
This stage also identifies relevant configuration parameters (e.g., model architecture or dataset inputs) that may be needed to allow for extensibility.

\subsubsection{Onboarding Funnel}

Using this record, candidate tools can be ``funneled'' and promoted through three stages: reproducibility, structural compatibility, and containerization.

\paragraph{Reproducibility check.}
This check determines whether the tool can be executed and whether it matches the target behavior reported by the original artifact. The candidate tool is run in an environment similar to the original reported setup (like runtime, software dependencies), using the same dataset, model, parameters, and evaluation metric. 
If the artifact fails due to minor dependencies or compatibility issues, the tool may still be considered replicable if minimal changes allow it to run without altering the core method \cite{Zhang2019TheoreticallyPT} (Refer to Table \ref{tbl:litSurveyFA}).

\paragraph{Structural check.}
If a tool is reproduced, the next phase determines its placement in the MLOps pipeline, what artifacts it consumes and produces according to the categories described in Section~\ref{subsec:pipeline-explorer}. 
This check records whether the tool is structurally composable in an MLOps pipeline. A tool may be composable as-is, may require a wrapper, or may be deemed non-composable. For instance, a tool that outputs outlier labels may become composable after a wrapper converts those labels into a filtered dataset that can then be transferred to other stages~\cite{Zhao2018XGBODIS}. In contrast, a tool that only reports a final score without producing a reusable artifact may be reproducible but not structurally composable\cite{Canonne2020TheDG} (Table \ref{tbl:litSurveyFA}).

The structural check uses the input/output of a particular tool to establish its relative positioning in the MLOps pipeline.
\landseer identifies the following \TML tool types:

\begin{itemize}
\item{\textbf{Pre-Training}}
techniques (\T{pre}{}) involve approaches that are applied before model training begins. Examples include dataset-level strategies that modify or transform the training data or methods that influence model architecture selection. The input and output here are datasets. 


\item{\textbf{During-Training}}
techniques (\T{in}{}) intervene directly in the learning algorithm. Examples include modifying the objective function, optimization process, model initialization, or gradient computations. Input is a dataset and model architecture, and output is a trained model.

\item{\textbf{Post-Training}}
techniques (\T{post}{}) are applied on a pre-trained model. While they do not modify the original training procedure, they still affect the internal state of the model. Examples include fine-tuning, weight pruning, or additional calibration steps. Input is a trained model and a dataset, and output is a modified model. 

\item{\textbf{Deployment}}
techniques (\T{dep}{}) 
are applied 
on the inputs to the model or its outputs during inference. Examples include input filtering to detect or block malicious queries, and output post-processing 
, and analyzing the model’s internal activations or behavior at inference time. Input is a model and a dataset, and output is a dataset. 

\end{itemize}

\paragraph{Containerization.}
This check determines whether the tool can be executed as an independent unit. Each tool passing the structural check is encapsulated in a standalone OCI~\cite{oci} image that bundles its dependencies and exposes a standard execution interface. 
The containerization phase also embeds the reproducibility record to aid Pipeline Explorer when scheduling experiments.
\subsection{Pipeline Explorer}
\label{subsec:pipeline-explorer}

Once tools are onboarded, the Pipeline Explorer is responsible for systematically evaluating how they behave in combination. 
It defines the experiment space, generates all valid defense compositions, schedules their execution, and coordinates between different runs.

The \landseer API utilizes two concepts to explore the composition space. First, an \textbf{Experiment}, which is defined by a model, a dataset, and a set of candidate defense tools for each stage of the ML pipeline. 
An experiment is a systematic exploration of all valid combinations of these tools, examining their collective effect on the model's utility and various \TML evaluation metrics, such as those listed in Table~\ref{tbl:evalmetric}. The results of all combinations are analyzed together to understand how different defenses interact and compare.
    
For each experiment, a series of \textbf{Combinations} are generated.
A combination is one specific combination of defense tools applied to a given model and dataset. 
It represents a single path through the experiment's search space, executing the selected tools in sequence across the ML stages and producing a set of evaluation metrics.



\subsubsection{Scheduler and Execution}

To carry out an experiment, one Scheduler and a series of Worker nodes (described below) operate in conjunction to carry out the following passes.

The scheduler proceeds in three steps. First, it \emph{validates} each tool 
against the experiment configuration, checking that the tool supports the 
requested dataset and that its input and output artifacts are compatible with 
adjacent stages. 
Second, it \emph{generates} all valid pipelines from the remaining 
tools, producing the full set of stage-wise ordered combinations. Third, it 
\emph{linearizes} each pipeline into an ordered sequence of atomic tasks suitable for distributed execution.

\paragraph{Scale and re-runs.}
Because many pipelines share common prefixes, naive re-execution is wasteful. 
The scheduler avoids redundant computation by assigning each task a content-derived identifier based on its tool, configuration, and input artifacts. 
If a matching result already exists in the cache, the task is skipped and its output is passed directly to the next stage. This deduplication significantly reduces total compute, particularly for Pre-Training and During-Training tasks that are shared across many pipelines. 
To support statistical robustness, the scheduler also allows pipelines to be re-run 
a configurable number of times with different random seeds, aggregating results across runs.

\paragraph{Workers} These are the execution units that carry out individual tasks. 
The scheduler assigns tasks to workers based on their resources (e.g., available GPUs), and workers execute tasks by invoking the corresponding \TML tool with the appropriate inputs.

\subsection{Interference Analysis}\label{sec:intanalysis}
\begin{figure*}[t] 
    \centering
    \includegraphics[trim=0 15 0 0, width=.99\linewidth]{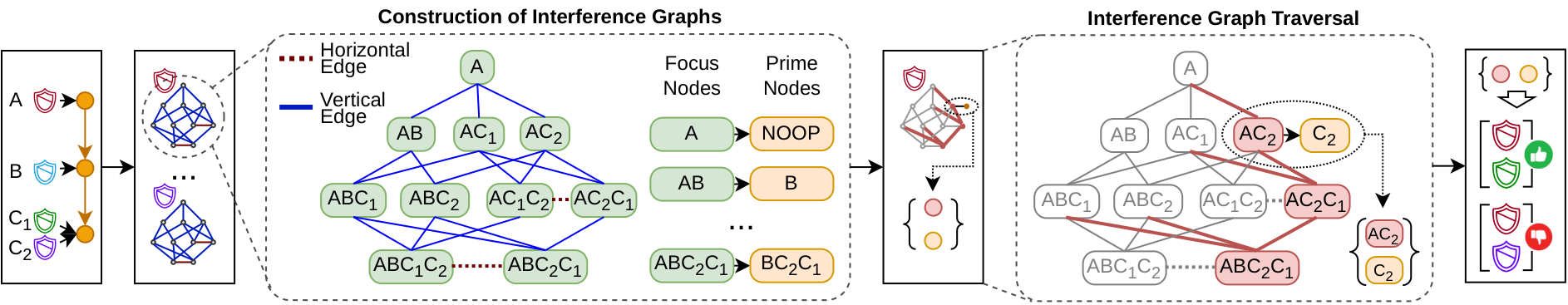}
    \caption{Results generation and analysis in \landseer. The pipeline result, ref Fig \ref{fig:overview}, is used as the input to the analysis algorithm, where comprehensive relationship graphs are built for each tool in the graph builder. A zoomed-in view of a sample comprehensive graph is seen in the Detailed Graph block. Each graph is passed through a traversal function, which draws different interference results.}
    \label{fig:interference_tree}
\end{figure*}

The results generated by \landseer (in Figure~\ref{fig:overview}) can be used to identify sources of interference.
\landseer presents these results as a vector $(C, M)$, where $C$ is a pipeline configuration (i.e., combination of tools) and $M=\{Acc, {m}_1, ..., {m}_n)$ is a list of performance criteria. Specifically, $Acc$ is the accuracy on clean test set, and $\{m_1,\cdots,m_n\}$ are $n$ trustworthy ML evaluation metrics.

We use performance metrics $M$ to assess qualitative composability of defenses. 
\landseer performs a root-cause analysis to identify the sources of interference (e.g., settings or techniques that trigger interference) within a flagged combination. Later, we introduce different categories of interference and present examples of defense combinations for each type. 

Given that \landseer generates a large quantity of combinations (our current corpus of tools generates more than \numcombinations different combinations) we must systematically identify only the ones that are indicative of interference.
To do so, we employ three steps for our data analysis pipeline:
\begin{inparaenum}
    \item build a data structure, namely an interference graph, that allows us to place any combination in relationship with other combinations,
    \item threshold interference by an adjustable value relative to a baseline measure, and
    \item classify interference patterns by inspecting the specifics between combination pairs and, optionally, their respective neighbors in the graph.
\end{inparaenum}

\subsubsection{Interference Graphs} We start by constructing \emph{interference graphs} for each defense tool (i.e., ``focus tool''). The nodes in the interference graph represent a combination of defense tools in the pipeline, with the root node consisting of a baseline combination $\textbf{c}_b$ with only the focus tool (and noop defenses in other stages). Similarly, other nodes in this inference graph will include the focus tool in combination with other defenses. Structurally, the interference graph holds a series of edges to represent \emph{minimum distances} between different combination nodes.
We describe graph edges as:

\begin{itemize}
    \item  Vertical edges: We depict incremental cardinality relationships between the number of tools in each combination (node) with a vertical edge. Child nodes add exactly one tool to the parent combination in the corresponding stage.
    Nodes that appear as parents for multiple top nodes are represented only once, with multiple vertical edges pointing to them to ensures a connected, non-redundant representation of all possible incremental tool combinations. 
    \item Horizontal edges: We define a neighboring combination as one in which the same tools are present but in different ordering, and depict neighboring combinations with a horizontal edge. 
    Keeping Horizontal relationships aid in identifying ordering-based interference as we discuss later.
\end{itemize}

Algorithm~\ref{alg:construction} describes the construction of $N$ graphs (one for each focus tool) from the set of all combinations $\mathcal{C}$ obtained from \landseer For each tool, we start from the lowest cardinality ($i=1$ in line 5) and extract the combination with only the focus tool $t$ (and noop in other stages) as the root node of the graph. In each following step, we extract nodes with cardinality $i$ which include tool $t$ ($FocusNodes$ in line 6) and add them to the graph. For each node, we also extract and save the \emph{prime} ($PrimeNodes$ in line 7) which has the same set of tools but without the focus tool. 
This focus-prime relationship is used for our interference analysis as we detail later.
For each node in the graph, we identify parent nodes (\texttt{Parent} function in line 12) and neighbor nodes (\texttt{Neighbor} function in line 17) and add vertical and horizontal edges respectively to the graph. 
As previously described, vertical edges connect parent nodes to child nodes which include all parent tools plus exactly one additional tool, preserving the order of tool application. Horizontal edges are added between nodes at the same cardinality level with the same tools in the combination but in a different order. This process continues until the graph $G_t$ for focus tool $t$ is completed. The previous steps are then repeated for the next focus tool until all $N$ graphs, namely the set of interference graphs $\mathcal{G}$ are constructed.

\begin{algorithm}[t]
    \caption{Construction of Interference Graphs}
    \label{alg:construction}
    \DontPrintSemicolon
    \KwIn{Set of all combinations $\mathcal{C}$, number of tools $N$}
    \KwResult{Set of interference graphs $\mathcal{G}$}
    \For{$t=1$ to $N$}{
        Initialize empty graph $G_t\gets \{\}$\;
        Initialize empty set $PrevNodes\gets \{\}$\;
        $k_t \gets \texttt{MaxCardinality}(\mathcal{C},t)$\;
        \For{$i=1$ to $k_t$}{
            $FocusNodes \gets \texttt{ExtractCombs}(\mathcal{C}, i, t)$\;
            $PrimeNodes \gets \texttt{ExtractPrime}(\mathcal{C}, FocusNodes)$\;
            $G_t.\texttt{Append}(FocusNodes)$\;
            $G_t.\texttt{Append}(PrimeNodes)$\;
            \If{$i>1$}{
                \For{$(n_1,n_2)$ in $(PrevNodes,FocusNodes)$}{
                    \If{$\texttt{Parent}(n_1,n_2)$}{
                        $G_t.\texttt{AddVerticalEdge}(n_1,n_2)$
                    }
                }
                \For{$(n_1,n_2)$ in $(FocusNodes,FocusNodes)$}{
                    \If{$\texttt{Neighbor}(n_1,n_2)$}{
                        $G_t.\texttt{AddHorizontalEdge}(n_1,n_2)$
                    }
                }
            }
            $PrevNodes \gets FocusNodes$
        }
        Add $G_t$ to $\mathcal{G}$\;
    }
    
\end{algorithm}

Figure~\ref{fig:interference_tree} depicts the construction of interference graphs for a pipeline with three stages with tools $A$ (``Pre-Training'' stage), $B$ (``During-Training'' stage), and $C1$, $C2$ (``Post-Training'' stage). The graph construction shows the resulting interference graph for tool $A$ as the focus tool. At each level of the tree, the child nodes are combinations with exactly one additional tool compared to their parents applied to its corresponding stage. 
Vertical edges are represented by blue solid lines and horizontal edges are represented by dotted red lines. 
For example, the line between nodes $AC_2C_1$ and $ABC_2C_1$ represent a parent-child relationship with tool $B$ added to the child node, whereas the line between nodes $AC_1C_2$ and $AC_2C_1$ represent a neighbor relationship with the order of tools $C_1$ and $C_2$ are reversed in the two combinations.

After constructing the interference graphs, we devise a technique to perform root-cause analysis of interference. 
As previously described in Section \ref{subsec:compo_defin}, we want to flag combinations where the addition of a focus tool degrades (or improves) performance across one of the metrics in $M$ compared to a baseline combination without the focus tool (prime combination).
Our root-cause analysis identifies combinations with interference that:
\begin{inparaenum}
    \item reduce (or increase) one or more of performance metrics in $M$ beyond a set threshold, and
    \item has the lowest cardinality that introduces such a drop.
\end{inparaenum}
This definition of root-cause combination ensures that the first instance of interference is identified by finding the lowers cardinality, and the interference is associated with the presence of the focus tool by comparing performance results to the prime node.

Our root-cause analysis is based on a graph traversal described in Algorithm~\ref{alg:traversal} on the interference graphs constructed for each tool. Starting from the highest cardinality nodes in the graph (line 3), we identify nodes with interference by comparing the performance metrics to the prime combination (\texttt{Interference} function in line 8). Any node with interference that has a parent node with interference is not considered root-cause, and the parent node is instead added for analysis next (lines 12-19).
The identified root-cause nodes together with their prime and neighbor nodes are then submitted to a characterization function described in Section~\ref{sec:intchar}. 

Figure~\ref{fig:interference_tree} illustrates the root-cause analysis steps through interference graph traversal. In this example, node $AC_2$ is identified as root-cause presenting interference while parent node $A$ and prime node $C_2$ do not show the same interference.  

\begin{algorithm}[t]
    \caption{Interference Graph Traversal}\label{alg:traversal}
    \DontPrintSemicolon
    \KwIn{Interference graph $G_t$}
    \KwResult{Set of identified combinations $\mathcal{S}$}
    Initialize empty set $\mathcal{S} \gets \{\}$\;
    Initialize empty set $NodeList \gets \{\}$\;
    $Nodes \gets \texttt{ExtractLeaves}(G_t)$\;
    $NodeList.\texttt{PushNode}(Nodes)$\;
    \While{$NodeList$ \text{not empty}}{
        $Node \gets NodeList.\texttt{PopNode}()$\;
        $PrimeNode\gets \texttt{ExtractPrime}(G_t, Node)$\;
        $NInt \gets \texttt{Interference}(Node, PrimeNode)$\;
        \If{$NInt$}{
            $RootCause \gets True$\;
            $ParentNodes \gets \texttt{ExtractParents}(G_t, Node)$\;
            \For{$PNode$ in $ParentNodes$}{
                $PrimeParent\gets \texttt{ExtractPrime}(G_t, PNode)$\;
                $PNInt \gets \texttt{Interference}(PNode, PrimeParent)$\;
                \If{$PNInt$}{
                    $NodeList.\texttt{PushNode}(PNode)$\;
                    $RootCause \gets False$\;
                }
            }
            \If{$RootCause$}{
                $Neighbors \gets \texttt{ExtractNeighbors}(G_t, Node)$\;
                $\mathcal{S}.\texttt{Add}([Node,PrimeNode,Neighbors])$\;
            }
        }
    }

\end{algorithm}

\noindent\textbf{Threshold Selection} To prevent random noise from affecting our analysis, we define threshold values and mark interference if the metric values change beyond these thresholds. We define two thresholds for each metric, namely $t_{l}$ and $t_h$. Changes to metric values below $t_l$ is classified as negligible, between $t_l$ and $t_h$ is classified as moderate, and above $t_h$ is classified as severe. We specify the values for $t_l$ and $t_h$ for our experiments in Section~\ref{exp_setup}.
This approach allows \landseer to classify the magnitude of metric changes, as well as the direction of the effect (positive, negative, or mixed), providing a standardized basis to evaluate when tool combinations meaningfully impact performance.


\subsubsection{Interference Characterization}\label{sec:intchar}
Having identified two $Comb$ nodes that are root-cause and above a threshold, we can characterize the type of interference. 
Intuitively, to characterize the type of interference, we label the combination by three properties:\begin{inparaenum}
    \item Cardinality difference between the nodes, as some sources of interference will appear on both vertical and horizontal edges (e.g., are the nodes on the same level or different levels?)
    \item Structural difference between the $Comb$ nodes (e.g., was there a change in ordering? was a new tool added in a particular stage of the pipeline?)
    \item Specific effect caused (e.g., did it affect accuracy? or a single TMLD value?)
\end{inparaenum}

This characterization can allow us to empirically label and categorize types of interference, under the following types: 



\begin{itemize}

    \item \textbf{Global Interference (GI)} indicates interference of a tool on others' performance metrics $M$ whenever it is applied in the combination.
    \item \textbf{Ordering Interference (OI)} represents interference between a set of tools only when a specific ordering of the tools is used.
    \item \textbf{Pairwise Interference (PW)} captures interference between exactly two tools as the change in their performance when used together vs. independently.

\end{itemize} 


\noindent We will use this classification in Section~\ref{sec:evalqual}.
\section{Implementation}

Beyond the architectural properties described above, \landseer's implementation required overcoming a series of challenges, primarily related to scale and combination pipelining.

\noindent\textbf{Challenge 1: Tool Dependency Management.}
Defenses within Landseer employ a variety of machine learning frameworks, Python versions, and CUDA toolkits. 
Artifacts identified frequently rely on a diverse set of dependency stacks, rendering a unified shared runtime impractical. 
Landseer encapsulates each defense in a self-contained runtime image with fixed dependencies and versioned environment specifications identified during onboarding.

Workers select and are selected from the available and supported backends on the host platform (e.g., Docker ~\cite{10.5555/2600239.2600241} for container-enabled systems or Apptainer/Singularity\cite{apptainer} for restricted HPC environments). 
The runtime contract remains consistent across different backends, ensuring the same mounted inputs, configuration interface, and output directory structure are utilized. 

\noindent\textbf{Challenge 2: Artifact Heterogeneity Across Pipeline Stages.}
Much like dependencies, intermediate artifacts use a variety of bespoke formats.
Every defense is scheduled with fixed mount points: \code{/data} (outputs from all upstream tasks), \code{/config} (model architecture and dataset parameters), and \code{/output} (where the tool writes its results). 
Landseer utilizes data stored in the reproduction record to identify and inject any necessary data representation transformation steps.
This may introduce concerns about the impact of such tasks; however, the onboarding process verifies reproducibility independently (i.e., outside of the framework) and in a standalone pipeline inside the framework that must match the target values in the reproduction record.

\noindent\textbf{Challenge 3: Exploring the Combinatorial Space.}
Running each workflow independently is computationally expensive because many workflows share common prefixes. 
For instance, workflows that start with the same preprocessing defense applied to the same dataset perform identical early-stage computations. 

\landseer compiles all valid combinations into a single global directed acyclic graph (DAG). 
Then, duplicate sub-paths are eliminated to avoid recomputation. 
Task equivalence is determined based on the tool used, combination configurations, and input artifact signatures. 
The scheduler processes the resulting graph in dependency order and prioritizes tasks likely to be reused. 
This approach enables shared prefixes to be computed once and reused across multiple downstream workflows, significantly improving computational efficiency.

\noindent\textbf{Challenge 4: Intermediate Artifact Caching and Data Management.}
Landseer employs a two-level artifact caching system. 
The first level consists of a local disk cache for each worker, enabling direct retrieval of artifacts previously produced on the same machine. 
The second level uses a shared MinIO~\cite{Minio} object store that is S3-compatible and accessible to all workers in the cluster.

When a worker completes a task, it stores the outputs locally and uploads them to the shared store. 
If another worker later needs the same artifact, it retrieves it from the shared store instead of re-executing the task. 
Much like other tools in this class, like CCache~\cite{ccache} \landseer's cached-artifacts are content-addressable, ensuring that equivalent identities across workers automatically reference the same stored artifact.
Local caches implement a least-recently-used (LRU) eviction policy once disk space usage exceeds a configurable threshold. In contrast, the shared store maintains complete artifact lineage for reproducibility and restartability.

\noindent\textbf{Challenge 5: Extensibility for Datasets, Models, Evaluators, and Tools}
\landseer is designed to support user-defined components without requiring modifications to the scheduler or execution engine. 
During the compilation of experiments, Landseer validates these specifications for schema correctness and stage compatibility. 
This plugin-style design allows for rapid expansion of benchmark coverage while ensuring reproducibility and compatibility.

\section{Evaluation}
\label{sec:evaluation}



%

In this Section, we study how \landseer can integrate \defenses into a pipeline, based on clearly defined criteria for structural composability, to then discover qualitative composability insights. 
We evaluate \landseer's ability to detect interference by both validating and refuting prior interference findings, while also identifying new types of interference based on empirical data. 

\subsection{Experimental Setup}
\label{exp_setup}

\paragraph{Datasets} Tools are tested on the CIFAR-10 dataset, a standard and widely used benchmark dataset for image classification tasks.

\paragraph{Model Architecture}
We use a CIFAR-style Residual Network (ResNet), ResNet-20, consisting of an initial 3×3 convolution followed by three stages of residual blocks with 16, 32, and 64 channels. 
The model has a fully connected classification layer, with weights initialized using Kaiming initialization \cite{Idelbayev18a}. 
For differentially private training using Opacus \cite{Abadi2016DeepLW}, batch normalization layers are configured with \textit{track\_running\_stats=False}. The modification avoids cross-batch information leakage through stored statistics and ensures that normalization depends only on the current mini-batch, which is more compatible with the per-sample gradient computations required by DP-SGD. 



\paragraph{Hyperparameters}
The baseline model is trained for 200 epochs with a batch size of 128 and an initial learning rate of 0.1. 
Standard CIFAR-10 data normalization was applied to both training and test sets. Data augmentation for training includes random cropping and random horizontal flipping. Hyperparameters for defenses tested in \landseer are listed in Table \ref{tbl:hyperparameters}.

For the qualitative composability studies, we set the metric thresholds $t_l$ and $t_h$ (from Section~\ref{sec:intanalysis}) to 2\% and 5\%, respectively.




\subsection{Structural Composability}
\label{sec:litreview}
\begin{table}[t]
\caption{Taxonomy of composability issues in practice and suggestions for improvement.}
\label{tbl:toolsfail}
\centering
\footnotesize
\setlength{\tabcolsep}{4pt}
\begin{tabular}{@{}p{0.075\textwidth} p{0.19\textwidth} p{0.18\textwidth}@{}}
\toprule

Composability class  &  Issues found in practice & Suggestions for improvement \\

\midrule
Available & 	\textasciitilde{30}\% lacked public artifacts; some only had third-party implementations ~\cite{Ciss2017ParsevalNI, Papernot2015DistillationAA, Zemel2013LearningFR, Borgnia2020StrongDA, Rouhani2019DeepSignsAE, zhang2018protecting}& Standardize official artifact release \\
\midrule

Reproducible & Software bugs, env. mismatch, missing configuration~\cite{Zhang2019TheoreticallyPT, Lin2023DifferentiallyPS, sablayrolles2020radioactive} & Provide validated code with clear instructions and requirements \\
\midrule

Replicable & Lack of end-to-end integration guidance~\cite{Canonne2020TheDG} & Provide minimal pipelines and integration examples \\
\midrule

Self-contained & All tools were containerizable, though with varying effort ~\cite{Li2023ReconstructiveNP}  & In addition to source code, provide ready-to-use container images where feasible \\

\midrule

Extensible & Some tools were limited by architecture-specific assumptions or requirements ~\cite{Abadi2016DeepLW,gu2017badnets} & Promote modular, model-agnostic designs where possible \\

\midrule

Modular & Cross-stage dependencies reduced modularity (e.g., requiring specific preprocessing and training choices) ~\cite{yao2022improving,gu2017badnets} &  Enforce stage separation where possible and document dependencies \\

\bottomrule
\end{tabular}

\end{table}
\begin{figure}[t]
  \centering
  \includegraphics[trim=0 40 0 0, width=\linewidth]{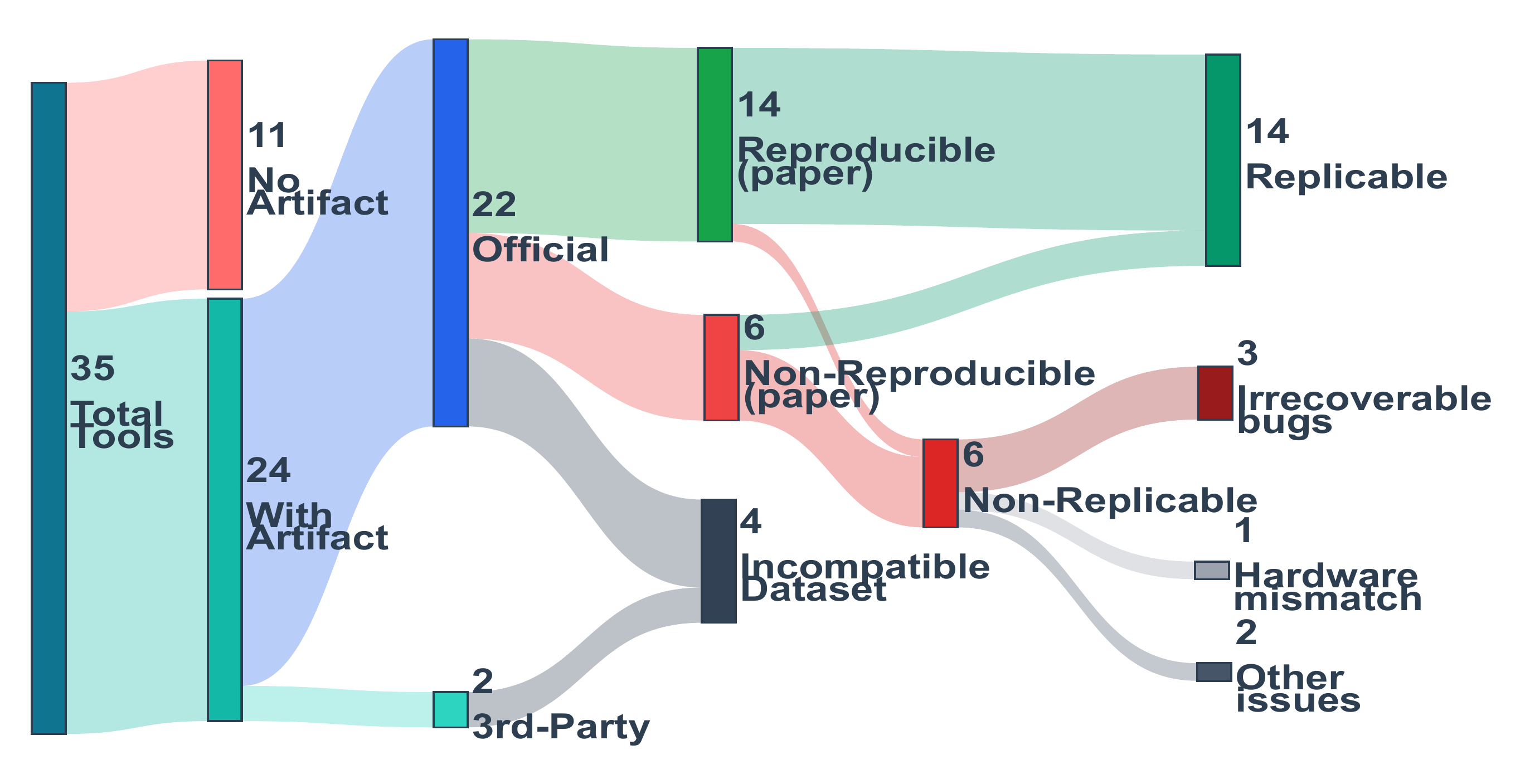}
  \caption{Summary of structural composability. Definitions for each category, as well as the flow and steps provided to assess composability are described in Section ~\ref{sec:litreview}}
  \label{fig:structuraloverview}
\end{figure}

We now empirically explore the landscape of \TML tools and study their structural composability. 
Thus, we first build a collection of tools from both industry and academia that address all the defense axes in TML (as reviewed in Section~\ref{sec:background}). Then, we study our ability to integrate them into an \mlops pipeline to identify the patterns and techniques that allow for structural composability.
As an added result, we also showcase the state of the tool ecosystem in terms of their composability. 




\subsubsection{\TML Tool Selection \& Integration Requirements}

We begin our composability study by collecting \TML tools from white and grey literature.
This includes academic publications in various venues for both machine learning and computer security in order to identify existing solutions.
Further, we enrich these solutions with bespoke systems from major open source machine learning frameworks (such as PyTorch). 
To ensure comprehensive and representative coverage, we survey work published over the past 12 years in top-tier conferences, journals, and other leading venues in machine learning, artificial intelligence, computer vision, and security, as summarized in Table \ref{tbl:papers}.
Specifically, we focus on identifying prior work addressing \TML tools for evasion robustness, outlier robustness, privacy, watermarking, fingerprinting, group fairness, and explanation. 
We select \TML tools to capture a representative sample that covers trustworthy tools across all ML axes and our structural categorization (as summarized in Table~\ref{tbl:litSurveyFA}). 

\subsubsection{Integration Barriers During Onboarding}
We define the structural composability criteria for tools and frameworks that implement each \TML defense and motivate them through analysis of studied tools. 
Each tool should feature an open-source implementation that delivers reproducible results. 
Also, the tool should integrate into an ML pipeline and work alongside other techniques. 

Specifically, we define the identify the following requirements for integration: 

\begin{itemize}
    \item \textit{Available}: Publicly available source code. This may be an official artifact from the authors or a third-party artifact.
    \item \textit{Reproducible}: Executable in a specific environment, and producing the reported results with the original code and data. 
    \item \textit{Replicable}: Executable in a specific environment with consistent results using different data and/or modified code.
    \item \textit{Self-contained}: Functional within an isolated environment (e.g., a Docker container \cite{10.5555/2600239.2600241}).
    \item \textit{Extensible}: Applicable to different datasets within a data modality (e.g., images) and model architectures.
    \item \textit{Modular}: It can be incorporated into one or more of the ML pipeline stages with described inputs and outputs to be passed into and out of its environment.
\end{itemize}


We treat these properties as forming a partially ordered hierarchy rather than a strictly linear one. 
Some requirements are inherently hierarchical; for example, an artifact must be available to be reproducible or replicable. 
Others are imposed by \landseer’s criteria: an artifact that is not self-contained is not considered structurally composable, even if it is otherwise extensible. 
Additionally, reproducibility and replicability are treated as being at the same level, since an artifact may not reproduce the original results exactly but can still yield consistent results under variation and thus remain usable within a pipeline (Note two non-reproducible papers marked as replicable on the top part of Figure~\ref{fig:structuraloverview}).

Defenses are categorized by their pipeline stage, but many are not \emph{stage-self-contained} within that stage. 
This makes it challenging to use them modularly in \landseer. 
A common issue is cross-stage coupling, where a method that claims to operate at one stage may actually produce a partially trained model, thereby dictating a fixed fine-tuning approach. 
That would mean During-Training, specific optimizers, loss functions, and schedules must be employed. 
This practice, which spans Pre-Training and During-Training, undermines modular assumptions. 
For instance, adversarial contrastive Pre-Training~\cite{jiang2020robust} necessitates particular downstream training decisions linked to its design.
A summary of additional barriers encountered includes:
\begin{itemize}
  \item \textbf{Non-standard I/O contracts:} bespoke formats, altered label spaces, or custom wrappers incompatible with generic pipelines;
  \item \textbf{Environment constraints:} reliance on legacy GPU capability or obsolete libraries;
  \item \textbf{Reproducibility gaps:} missing code, seeds, or configuration details.
\end{itemize}

These barriers, more of which are listed in Table \ref{tbl:toolsfail}, underscore the need for explicit \emph{structural composability} criteria in selecting defenses for integration.

For each defense with publicly available artifacts, we first attempted to reproduce the results reported in the original paper. 
To assess reproducibility, we define an error tolerance that accounts for expected variability in machine learning pipelines. 
In particular, we consider a defense reproducible if its performance falls within a small deviation threshold of the reported results. 
We set this threshold to 3\%, which reflects typical run-to-run variation observed in practice while remaining strict enough to detect substantive discrepancies. 
Minor deviations of this magnitude can arise from nondeterminism in GPU operations, floating-point precision differences across hardware, subtle changes in library versions, or undocumented preprocessing steps. 



Some defenses, especially older ones, do not execute properly in modern machine learning toolchains due to deprecated APIs or outdated dependencies. 
For those cases, we modified the implementations as little as possible to ensure compatibility, without altering their core logic or methodology. 
To verify that these modifications did not affect performance, we re-ran the original experiments and confirmed that the results deviated from the published values by no more than 3\%. 

Further, many defenses are implemented with author-chosen optimizers and learning-rate schedulers (e.g., SGD, Adam, AdamW, DP-SGD). 
This variation can complicate comparisons because differences in utility and robustness may arise from the optimizer or schedule rather than the defense itself. 
In this iteration of Landseer, we maintain the original optimizer and schedule from each paper to ensure accuracy. 
Currently, if two tools have different optimizers and this affects utility or defense, we will consider them as negatively interfering and hence not composable.
However, we also acknowledge that systematically varying these factors is crucial for exploring the interplay between optimizers and defenses, and their composability. 

Lastly, given that some solutions are intended more as technology demonstrations rather than components to be directly integrated into a pipeline, we make minimal modifications to their implementations to allow for chaining of steps. 
For example, in \cite{Zhao2018XGBODIS}, the defense code outputs an array of 0s and 1s to indicate whether a data point is an outlier (1) or not (0), but this output cannot be directly passed to the next step in the pipeline. 
To address this, we instruct \landseer to add a post-processing stage to remove outliers before passing the resulting artifact forward. 

\subsubsection{Structural Landscape}
\label{subsec:rq1.3}

We present the results of the onboarding process in Figure \ref{fig:overview}. 
From a total of \numselectedtools original papers, \numnoartifacts were excluded due to the absence of public artifacts. 
Of the remaining \numwithartifacts papers, \numofficialartifacts had official artifacts and \numthirdparty third-party implementations. 
All \numthirdparty third-party implementations, along with 2 more official implementations, were dropped as they focused on tabular datasets, while Landseer focused on image datasets. 
Of the remaining 20 official artifacts, 14 were reproducible and replicable, and the remaining 6 were dropped due to irrecoverable bugs that made the artifacts fail to run.
Of 14 replicable papers, some were included directly into \landseer without changes while the rest required minimal modifications, such as updating deprecated packages, updating python versions, and updating API function calls.

\subsection{Qualitative Composability}\label{sec:evalqual}




We follow our structural composability evaluation with conducting qualitative evaluations to demonstrate \landseer capabilities. Specifically, we focus on two case studies. In the first case study, we compare \landseer results to existing work on pairwise defense combinations to evaluate if \landseer can reproduce prior conjectures and examine any discrepancies. In the second case study, we explore new combinations that are identified by \landseer with interesting properties that have not been examined in prior work, demonstrating the extensibility of \landseer.

\subsubsection{Case Study 1: Comparison to Existing Pairwise Defense Combinations}
\newcolumntype{L}{>{\raggedright\arraybackslash}X}

\definecolor{alignedbg}{HTML}{E3F2FD}
\definecolor{alignedfg}{HTML}{1976D2}
\definecolor{conflictingbg}{HTML}{FDECEA}
\definecolor{conflictingfg}{HTML}{C62828}
\definecolor{egoodfg}{HTML}{0D47A1}
\definecolor{ebadfg}{HTML}{B71C1C}

\newcommand{\StatusGood}{\colorbox{alignedbg}{\strut\textcolor{alignedfg}{\textbf{No}}}}
\newcommand{\StatusBad }{\colorbox{conflictingbg}{\strut\textcolor{conflictingfg}{\textbf{Yes}}}}

\newcommand{\egood}{\textcolor{egoodfg}{\LARGE$\bullet$}}
\newcommand{\ebad}{\textcolor{ebadfg}{\LARGE$\bullet$}}
\newcommand{\drgood}{\textcolor{egoodfg}{$\blacksquare$}}
\newcommand{\drbad}{\textcolor{ebadfg}{$\blacksquare$}}
\newcommand{\none}{\textcolor{lightgray}{\LARGE$\bullet$}} 

\begin{table}[h!]
\footnotesize
\renewcommand{\arraystretch}{1.15}
\caption{Comparison of \landseer findings on pairwise tool combinations with existing work. We indicate \landseer confirming (contradicting) prior work with \egood{} (\ebad{}).}


\begin{tabular}{p{1.6cm}cp{2.0cm}p{2.0cm}}
\toprule

\textbf{Tool Category} & \textbf{Conflicting} & \multicolumn{2}{c}{\textbf{Existing Work}} \\
\cmidrule(lr){3-4}

 & & \textbf{Confirm} & \textbf{Contradict} \\
 
\midrule



\AR{in}{}   / \OR{post}{} & \StatusBad &  \ebad \;\cite{duddu2024combining} & \none \\

\midrule

\AR{in}{} / \FP{dep}{}  & \StatusBad & \ebad \;\cite{Lukas2019DeepNN} & \egood \;\cite{szyller2023conflicting, duddu2024combining}  \\

\midrule 

\WM{pre}{} / \AR{in}{} & \StatusBad &  \ebad \;\cite{duddu2024combining, szyller2023conflicting} & \none \\

\midrule

\AR{in}{} / \EX{dep}{} & \StatusBad & \none & \egood \;\cite{duddu2024combining} \\

\midrule

\OR{post}{} / \FP{dep}{} & \StatusGood & \egood \;\cite{duddu2024combining} & \none \\

\midrule

\WM{pre}{} / \OR{post}{} & \StatusBad & \ebad \;\cite{duddu2024combining} & \none \\

\midrule

\OR{post}{} / \EX{dep}{} & \StatusGood & \egood \;\cite{duddu2024combining} & \none \\

\midrule

\FP{dep}{} / \EX{dep}{} & \StatusGood & \egood \;\cite{duddu2024combining} & \none \\

\midrule 

\WM{pre}{} / \EX{dep}{}  & \StatusGood & \egood \;\cite{duddu2024combining} & \none \\

\midrule
\OR{pre}{} / \DP{in}{} & \StatusGood & \egood \;\cite{9802763} & \none \\

\midrule
\AR{in}{} / \DP{dep}{} & \StatusBad & \ebad \;\cite{9802763, DBLP:journals/corr/abs-2201-02265} & \none \\

\midrule 
\DP{in}{} / \AR{post}{} & \StatusBad & \ebad \;\cite{DBLP:journals/corr/abs-2201-02265} & \none \\

\midrule
\DP{in}{} / \FP{dep}{} & \StatusBad & \none & \egood \;\cite{duddu2024combining, szyller2023conflicting} \\

\midrule

\DP{in}{} / \EX{dep}{} & \StatusGood & \egood \;\cite{duddu2024combining} & \none \\

\midrule
\DP{in}{} / \OR{post}{} & \StatusGood & \egood \;\cite{9802763} & \none \\













\bottomrule

\end{tabular}

\label{tab:pairwise_comparison}
\end{table}

We begin by evaluating \landseer's interference results against existing work on interactions between defenses and summarize our results in Table~\ref{tab:pairwise_comparison}. In this table, we define and mark conflicting combinations as one in which one or more performance metrics degrade beyond the thresholds defined in Section \ref{exp_setup}, and not conflicting (aligning) as one with improved or unchanged metrics. 
To demonstrate the ability of \landseer to reproduce existing results, we perform a case study with five papers in literature by Duddu et al.~\cite{duddu2024combining}, Szyller et al.~\cite{szyller2023conflicting}, Lukas et al. ~\cite{Lukas2019DeepNN}, Hayes et al. ~\cite{DBLP:journals/corr/abs-2201-02265}, and Strobel et al. ~\cite{9802763} that examine interactions of pairwise defenses.
In particular, Duddu et al. ~\cite{duddu2024combining} propose a non-experimental prediction technique, DEF\textbackslash{}CON, that uses a tool questioning framework to explore interactions between pairwise defenses to determine if a combination of defenses will be aligning, and Szyller et al. ~\cite{szyller2023conflicting} empirically examined the pairwise combination of model/data ownership verification with differentially private training and model evasion robustness, to find aligning or conflicting interactions.
%
For each combination, we inspect
when \landseer confirms or contradicts the results in prior work.
As shown in Table~\ref{tab:pairwise_comparison}, \landseer obtains the same results for 13 different pairwise defense combinations, confirming the existence (lack) of interference in 8 (7) cases.

%
%
%
%
%

\noindent\textbf{Confirmations.}
Duddu et al. \cite{duddu2024combining} determine that model Explanation techniques applied after model training \EX{dep}{} do not conflict with watermarking \WM{pre}{}, differential privacy \DP{in}{}, poisoning robustness \OR{post}{}, or fingerprinting \FP{dep}{}.
On the other hand, they report interference between watermarking applied before training \WM{pre}{} and robustness defenses (adversarial robustness \AR{in}{} and poisoning robustness \OR{post}{}). Similarly, their results show interference between adversarial robustness \AR{in}{} and backdoor removal \OR{post}{}. \landseer confirms these results, sharing four of the six tools they use in their evaluation, and differing in two. The two that differ are \OR{post}{} where they use a technique by Zheng et al. ~\cite{Zheng_pre_activation}, and \EX{dep}{} where they use DeepLift \cite{deeplift}.
Interestingly, Duddu et al. initially predict that the \AR{in}{}-\OR{post}{} combination does not result in interference based on their proposed model. However, their empirical results confirm the existence of interference, demonstrating the challenges of prediction models for defense interactions. 



Szyller et al. \cite{szyller2023conflicting} confirm the interference between watermarking \WM{pre}{} and adversarial robustness \AR{in}{}, since adversarial training suppresses introduced watermarks in most cases. Hayes et al. ~\cite{DBLP:journals/corr/abs-2201-02265} and Strobel et al. ~\cite{9802763} determine that adversarial robustness \AR{in}{} and differential privacy \DP{dep}{} are interfering. Strobel et al. claim that adversarial training, a form of adversarial robustness technique, modifies model decision boundaries in a way that an attacker can exploit to craft successful inference attacks, and Hayes et al. claim that adversarial perturbations and clipping norms lead to poor model generalization.


\noindent\textbf{Contradictions.}
Duddu et al. and Szyller et al. report no interference between adversarial robustness \AR{in}{} and fingerprinting \FP{dep}{}, whereas \landseer empirical data (using the same techniques: \AR{in}{} ~\cite{Zhang2019TheoreticallyPT} and \FP{dep}{} ~\cite{Maini2021DatasetIO}) contradicts this finding. We note that Lukas et al.~\cite{Lukas2019DeepNN}, which also investigates this combination with different \AR{in}{} and \FP{dep}{} techniques on CIFAR10 and Resnet20, claim a conflicting interaction between the two tools.
Duddu et al. report their result based on their prediction model and without performing any empirical evaluation for this setting.
Szyller et al. \cite{szyller2023conflicting} conduct experiments using a different adversarial robustness technique ~\cite{Madry2017TowardsDL} but only report metrics for fingerprinting and do not include metrics for the adversarial robustness tool to confirm their claim. While the absence of this data does not necessarily invalidate their claim, the use of a different adversarial robustness technique could explain the discrepancy in results. Interference effects may depend on the specific \TML techniques used, an aspect that future extensions of \landseer aim to investigate further.


In a new pairwise combination between differential privacy \DP{in}{} and fingerprinting \FP{dep}{}, both Duddu et al. and Szyller et al. claim an aligning combination, where Duddu et al. rely solely on their prediction model while Szyller et al. use empirical data to show fingerprinting is not affected, but not the value for differential privacy. Additionally, discrepancies with \landseer may stem from differences in experimental settings, including datasets, model architectures, and evaluation metrics. While overlapping components (e.g., CIFAR-10 and ResNet-20) are used, the exact model–dataset mappings in Szyller et al. are unclear. Moreover, they evaluate differential privacy using classification accuracy, whereas \landseer uses membership inference attack AUC, making comparison difficult.

In the case of adversarial robustness and explanation techniques, there is a contradiction between \landseer's results and \cite{duddu2024combining}. Duddu et al. \cite{duddu2024combining} claim that there is no interference between adversarial robustness \AR{in}{} and explanation \EX{dep}{}, whereas \landseer discovered interference mainly on the explanation defense. The difference in interference claims could be attributed to the use of different explanation techniques and experimental setups. \cite{duddu2024combining} used the DeepLift \cite{deeplift} technique and tested on FMNIST and UKTFACE with a 2-layer CNN model, whereas \landseer implemented DeepSHAP \cite{ExplainShap_NIPS2017_8a20a862} on CIFAR10 and a Resnet20 model.


\begin{table}[t]
\footnotesize
\centering
\caption{Subset of new defense combinations with interference state. Interference will be indicated as follows: Pairwise (PW), Ordering (OI), and Global (GI). 
For each metric, $\uparrow$ ($\downarrow$) indicates improved (degraded), + + (- -) represents severe improvement (degradation), + (-) refers to moderate improvement (degradation), and $\equiv$ denotes no change. A bidirectional arrow $\Leftrightarrow$ indicates that swapping the order of tools in the same defense stage will observe the metric changes recorded.}
\label{tbl:interference_summary}

\renewcommand{\arraystretch}{.5} 
\begin{tabular}{ll p{2.0cm}}
\toprule
\textbf{Combination} & \textbf{Interf.} & \textbf{Metric Change} \\

\midrule

\OR{pre}{} / \AR{in}{} & PW & $\textbf{m}^\textbf{ar}\uparrow$++, $\textbf{m}^\textbf{ev} \uparrow$++\\

\midrule
\AR{post}{} / \FP{dep}{} & PW & $\textbf{m}^\textbf{ev} \equiv$, $\textbf{m}^\textbf{fp} \equiv$ \\

\midrule
\OR{pre}{}+\WM{pre}{} & PW & $\textbf{m}^\textbf{ar} \equiv$, $\textbf{m}^\textbf{wm} \equiv$ \\  
\midrule
\AR{in}{} / \AR{post}{} & PW & $\textbf{m}^\textbf{ev} \uparrow$++ \\

\midrule
\AR{post}{} / \DP{dep}{} & PW & $\textbf{m}^\textbf{ev} \downarrow$- -,$\textbf{m}^\textbf{dp}_\mathbf{mia} \equiv$ \\
\midrule 
\OR{pre}{} / \EX{dep}{} & PW & $\textbf{m}^\textbf{ar} \equiv$, $\textbf{m}^\textbf{ex} \downarrow$- - \\
\midrule
\WM{pre}{}+\OR{pre}{} / \AR{in}{} / \AR{post}{}+\OR{post}{} & GI & $\textbf{m}^\textbf{wm} \equiv$, $\textbf{m}^\textbf{ar} \equiv$, $\textbf{m}^\textbf{ev} \downarrow$- - , $\textbf{m}^\textbf{ou} \equiv$ \\
\midrule
\WM{pre}{}+\OR{pre}{} / \AR{in}{} / \AR{post}{}+\OR{post}{} / \FP{dep}{} & GI & $\textbf{m}^\textbf{wm} \equiv$, $\textbf{m}^\textbf{ar} \equiv$, $\textbf{m}^\textbf{ev} \uparrow$++, $\textbf{m}^\textbf{ou} \equiv$ \\

\midrule
\WM{pre}{}+\OR{pre}{} / \AR{in}{} / \AR{post}{}+\OR{post}{} / \EX{dep}{} & GI & $\textbf{m}^\textbf{wm} \equiv$, $\textbf{m}^\textbf{ar} \equiv$, $\textbf{m}^\textbf{ev} \uparrow$++, $\textbf{m}^\textbf{ou} \equiv$ \\

\midrule
\OR{pre}{} $\Leftrightarrow$ \WM{pre}{} / \OR{post}{} & OI & $\textbf{m}^\textbf{ar} \downarrow$- , $\textbf{m}^\textbf{wm} \uparrow$++ \\

\midrule
\OR{pre}{} $\Leftrightarrow$ \WM{pre}{} / \AR{in}{} & OI & $\textbf{m}^\textbf{ar} \equiv$ , $\textbf{m}^\textbf{wm} \uparrow$++ \\

\midrule
\WM{pre}{} $\Leftrightarrow$ \OR{pre}{} / \AR{post}{}+\OR{post}{} & OI & $\textbf{m}^\textbf{ar} \uparrow$+ , $\textbf{m}^\textbf{wm} \downarrow$- - \\

\bottomrule
\end{tabular}

\end{table}

\subsubsection{Case Study 2: New Defense Combinations}
In our next case study, we examine new defense combinations identified by \landseer that, to the best of our knowledge, have not been explored in prior work. We show that overlooked factors such as the stage at which a defense is applied or the order of defenses in the same stage can result in different interference patterns.

\Paragraph{Unexplored Pairwise Combinations.}
In Table \ref{tbl:interference_summary} we show examples of pairwise defense combination that have not been explored previously but prove to be insightful. Specifically, \landseer discovers that incorporating outlier removal in the Pre-Training stage \OR{pre}{} and adversarial training \AR{in}{} significantly improve the corresponding metrics. This result is noteworthy since prior conclusions by Duddu et al. (also confirmed by \landseer) indicate that applying outlier robustness techniques Post-Training \OR{post}{} results in a conflicting combination with \AR{in}{} (Table~\ref{tab:pairwise_comparison}).
The same trend is seen where a previously explored combination of adversarial training \AR{in}{} and fingerprinting \FP{dep}{} may give conflicting results (Table~\ref{tab:pairwise_comparison}), but applying the adversarial robustness technique in the post stage \AR{post}{} could be a better option for aligning the defenses as our new results indicate in Table~\ref{tbl:interference_summary}.

\landseer reveals another interesting pairwise combination where applying watermaking \WM{pre}{} and outlier removal \OR{pre}{} (both in the Pre-Training stage) does not degrade either metric, while previously explored 
combinations of watermaking \WM{pre}{} and outlier robustness \OR{post}{} observed a conflicting combination (Table \ref{tab:pairwise_comparison}).

For defenses with the same objective, we further notice that their pairwise combination could be beneficial as a layered defense strategy.
For example, Table \ref{tbl:interference_summary} shows that applying adversarial robustness techniques to During-Training \AR{in}{} and Post-Training \AR{post}{} results in a significant improvement in the adversarial robustness metric 
compared to combinations where only one is present.

Other previously unexplored pairwise combinations in Table~\ref{tbl:interference_summary}, namely \AR{post}{}/\DP{dep}{} and \OR{pre}{}/\EX{dep}{}), indicate interference with degraded metric values. Comparing this insight with prior results in Table~\ref{tab:pairwise_comparison} with the same defense in other stages reveals agreement in one case (conflicting \AR{in}{}/\DP{dep}{}) and disagreement in the other case (aligning \OR{post}{}/\EX{dep}{}), hinting at other factors that may be at play here worth exploring in future work.

\Paragraph{Combinations with Global Interference} 
\landseer is also able to explore combinations of 2+ tools which have been mostly overlooked in prior work. 
Our results reveal clear stage-dependent patterns. For example, Pre-Training tools (\WM{pre}{}, \OR{pre}{}) and During-Training (\DP{in}{},\AR{in}{}) tools consistently show interference when combined with other tools (>95\% of combinations). This result indicates that defenses applied at later stages may have better composability with other techniques which can be an interesting avenue for future work.
%
%
We include a few examples of combinations with 5+ tools in Table~\ref{tbl:interference_summary} that show interesting global interference trends (marked with GI). The three focus tools here, namely \EX{dep}{}, \FP{dep}{}, and \AR{post}{}, do not exhibit global interference except in the specific arrangement shown in the table. Notably, the addition of \FP{dep}{} and \EX{dep}{} improve the robustness metric.
More detailed exploration of these combinations could reveal more insights about the interactions of defenses when several coexist in the pipeline.


\Paragraph{Combinations with Ordering Interference} 
Finally, we present our results for ordering interference (marked by OI in Table~\ref{tbl:interference_summary}). Notably, applying \OR{pre}{} and \WM{pre}{} alone does not result in ordering interference. However, 
%
in the presence of \OR{post}{} or \AR{dur}{}, the ordering between \OR{pre}{} and \WM{pre}{} becomes important, with \OR{pre}{} before \WM{pre}{} producing more favorable outcomes.

\section{Discussion}


\noindent \landseer sits at the crossroads of many efforts to improve \defenses in the \mlops context.
Naturally, this suggests there exist various efforts and approaches to improve the status quo. 
Particularly, work on replicability and composability has been broached in various ways, and exploring their overlap --- beyond what a typical related work section would --- is paramount.
Lastly, we also explore ways in which \landseer can be improved or modified.

\Paragraph{Contemporary Work on Reproducibility.} Existing work has explored the way in which \defenses papers are reproducible in practice.
Notably, Olszewski et al.~\cite{Olszewskireproducibility2023} carried out a large-scale replication experiment similar to our structural reproducibility work.
Our results complement some of the findings in that work, in that we found similar patterns and challenges during our replication phase --- they found that about 20\% of papers were replicable, while we found that about 40\% were so.
However, \landseer expands this notion by also providing visibility on how these papers can integrate into an \mlops pipeline.

\Paragraph{Contemporary Work on Composability.} Similarly, there has been an uptick in work relating to our notions of \defenses' composability.
In this regard, contemporary work (like \cite{duddu2024combining,nagireddy2023functioncompositiontrustworthymachine, 9797363}) mostly provides a manual analysis approach to explore piecemeal combinations of tools.
Instead, \landseer provides a systematic framework that: 
    1) Allows researchers to quickly test the composability of novel systems by packaging their \defenses as a module.
    2) Allows practitioners to test a specific pipeline configuration to find sources of interference
    3) Provides guidance on defense alternatives of pipeline configurations that minimize interference


\Paragraph{Tuning \defenses Parameters to Improve Composability.} An intuitive notion to improve composability may be by tweaking \defenses individual parameters.
For example, by reducing the $\sigma$ for a differential privacy defense, we may reduce its interference on other \defenses.
However, searching for optimal operations and determining acceptable thresholds would require substantial changes to the way that solutions are composed today, and how the \sys pipeline executor explores what effectively is an n-dimensional hyperspace.
We defer these extensions to \sys to future work as a consequence.

\section{Related Work}




Much existing literature focuses on single defenses for the known threats in the ML/AI supply chain such as adversarial robustness for adversarial attacks \cite{Cohen2021SimplePR, Shafahi2019AdversarialTF, Zheng2019EfficientAT, Yang2020RandomizedSO}, outlier removal and pruning for poisoning and backdoor attacks \cite{Wu2021AdversarialNP, Hong2020OnTE, Li2021NeuralAD, yang2022not}, among others \cite{Adi2018TurningYW, Lukas2021SoKHR, Subramanian2023HaveTC, Ross2017RightFT, Lin2023DifferentiallyPS, Peng2022FingerprintingDN} as discussed in section 3. Multiple defenses are needed to properly secure ML models and AI systems. 

Recent work focused on combining at most two defenses and studying the interactions/conflicts between those defenses \cite{gittens2022adversarial, chen2023privacy,noppel2024sok}. Szyller et al. \cite{szyller2023conflicting} study pairwise combination of ownership verification with differential privacy and evasion robustness, finding frequent conflicts. Duddu et al. \cite{duddu2024sok} show that defenses targeting one risk (security, privacy, fairness) can unintentionally affect others.  In \cite{duddu2024combining}, Duddu et al. propose a non-experimental framework to assess defense compatibility. It is limited to two defenses and has little validation for larger combinations. Cuong Tran et al. \cite{tran2024effects} study trade-offs between evasion robustness and fairness, proposing a framework for systems with good trade-offs between robustness and fairness. Additionally, work by Khan et al. \cite{hassanpour2024impact} highlights the complexity of multi-objective defense design, showing that improving one objective (e.g., privacy via regularization) can degrade utility or fairness in image classification, reinforcing the need for careful composition of defenses.

Despite significant efforts in the related work to address the challenge of combining ML defenses, there is still no infrastructure that enables the seamless composition of multiple ML defenses with empirical guarantees. Existing combination techniques are limited to two defenses at a time, and there is no systematic way to evaluate the effects of adding more defenses or tuning their parameters within a unified framework.

\section{Conclusion \& Future Work}

This paper introduces \landseer, a framework to systematically explore \TML tool composability.
\landseer can easily integrate a defense in reference machine learning pipelines, and compute an exhaustive combination of all configurations with all other tools.
Afterwards, it identifies their interference and catalogs it in a pipeline-oriented taxonomy.
Using \landseer we were able to identify various sources of \defenses interferences, as well as contrast with contemporary work.

Though the current implementation of \landseer is relatively coarse, for it does not allow exploring low-level semantics that may influence interference, such as optimizers, batching approach, \defenses security parameters, etc., we envision future extensions of the framework will allow us to provide insight into these.
Similarly, the inclusion of a larger corpus of work may also allow us to improve the insights we derive from our interference function.
However, as is, \landseer is already able to provide useful insights for researchers and practitioners around ML composability.

\bibliographystyle{ACM-Reference-Format}
\bibliography{references}

@article{petsiuk2018rise,
  title={Rise: Randomized input sampling for explanation of black-box models},
  author={Petsiuk, Vitali and Das, Abir and Saenko, Kate},
  journal={arXiv preprint arXiv:1806.07421},
  year={2018}
}

@article{tsai2023adversarial,
  title={Adversarial attacks on medical image classification},
  author={Tsai, Min-Jen and Lin, Ping-Yi and Lee, Ming-En},
  journal={Cancers},
  volume={15},
  number={17},
  pages={4229},
  year={2023},
  publisher={MDPI}
}

@article{Borgnia2020StrongDA,
  title={Strong Data Augmentation Sanitizes Poisoning and Backdoor Attacks Without an Accuracy Tradeoff},
  author={Eitan Borgnia and Valeriia Cherepanova and Liam H. Fowl and Amin Ghiasi and Jonas Geiping and Micah Goldblum and Tom Goldstein and Arjun Gupta},
  journal={ICASSP 2021 - 2021 IEEE International Conference on Acoustics, Speech and Signal Processing (ICASSP)},
  year={2020},
  pages={3855-3859},
  url={https://api.semanticscholar.org/CorpusID:227054251}
}

@InProceedings{Zhang_2024_CVPR,
    author    = {Zhang, Jinghuai and Liu, Hongbin and Jia, Jinyuan and Gong, Neil Zhenqiang},
    title     = {Data Poisoning based Backdoor Attacks to Contrastive Learning},
    booktitle = {Proceedings of the IEEE/CVF Conference on Computer Vision and Pattern Recognition (CVPR)},
    month     = {June},
    year      = {2024},
    pages     = {24357-24366}
}

@inproceedings{Zheng_pre_activation,
author = {Zheng, Runkai and Tang, Rongjun and Li, Jianze and Liu, Li},
title = {Pre-activation distributions expose backdoor neurons},
year = {2022},
isbn = {9781713871088},
publisher = {Curran Associates Inc.},
address = {Red Hook, NY, USA},
booktitle = {Proceedings of the 36th International Conference on Neural Information Processing Systems},
articleno = {1356},
numpages = {14},
location = {New Orleans, LA, USA},
series = {NIPS '22}
}

@inproceedings{deeplift,
author = {Shrikumar, Avanti and Greenside, Peyton and Kundaje, Anshul},
title = {Learning important features through propagating activation differences},
year = {2017},
publisher = {JMLR.org},
pages = {3145–3153},
numpages = {9},
location = {Sydney, NSW, Australia},
series = {ICML'17}
}

@inproceedings{ExplainShap_NIPS2017_8a20a862,
 author = {Lundberg, Scott M and Lee, Su-In},
 booktitle = {Advances in Neural Information Processing Systems},
 editor = {I. Guyon and U. Von Luxburg and S. Bengio and H. Wallach and R. Fergus and S. Vishwanathan and R. Garnett},
 pages = {},
 publisher = {Curran Associates, Inc.},
 title = {A Unified Approach to Interpreting Model Predictions},
 url = {https://proceedings.neurips.cc/paper_files/paper/2017/file/8a20a8621978632d76c43dfd28b67767-Paper.pdf},
 volume = {30},
 year = {2017}
}

@misc{Idelbayev18a,
  author       = "Yerlan Idelbayev",
  title        = "Proper {ResNet} Implementation for {CIFAR10/CIFAR100} in {PyTorch}",
  howpublished = "\url{https://github.com/akamaster/pytorch_resnet_cifar10}",
  note         = "Accessed: 2025-05-06"
}

@article{10.5555/2600239.2600241,
author = {Merkel, Dirk},
title = {Docker: lightweight Linux containers for consistent development and deployment},
year = {2014},
issue_date = {March 2014},
publisher = {Belltown Media},
address = {Houston, TX},
volume = {2014},
number = {239},
issn = {1075-3583},
journal = {Linux J.},
month = mar,
articleno = {2}
}

@misc{apptainer,
author={Apptainer},
title={Apptainer},
url={https://apptainer.org/},
journal={Apptainer.org},
year={2023} }

@misc{Minio,
    title={{high performance data store for AI \& Analytics  minio\_2026}},
    url={https://www.min.io/},
    author="{The MinIO project}",
    journal={Www.min.io},
    year={2026} 
}

@misc{ccache,
    title={{ccache — compiler cache\_2026}},
    url={https://ccache.dev/},
    author="{CCache Developers}",
    journal={Ccache.dev},
    year={2026} 
}

@misc{oci,
  author       = {{Open Container Initiative}},
  title        = {Open Container Initiative},
  year         = {2015},
  howpublished = {\url{https://opencontainers.org}},
}

@misc{agarwal2018reductionsapproachfairclassification,
      title={A Reductions Approach to Fair Classification}, 
      author={Alekh Agarwal and Alina Beygelzimer and Miroslav Dudík and John Langford and Hanna Wallach},
      year={2018},
      eprint={1803.02453},
      archivePrefix={arXiv},
      primaryClass={cs.LG},
      url={https://arxiv.org/abs/1803.02453}, 
}

@article{pessach2022review,
  title={A review on fairness in machine learning},
  author={Pessach, Dana and Shmueli, Erez},
  journal={ACM Computing Surveys (CSUR)},
  volume={55},
  number={3},
  pages={1--44},
  year={2022},
  publisher={ACM New York, NY}
}

@INPROCEEDINGS{8406613,
  author={Papernot, Nicolas and McDaniel, Patrick and Sinha, Arunesh and Wellman, Michael P.},
  booktitle={2018 IEEE European Symposium on Security and Privacy (EuroS\&P)}, 
  title={SoK: Security and Privacy in Machine Learning}, 
  year={2018},
  volume={},
  number={},
  pages={399-414},
  doi={10.1109/EuroSP.2018.00035}}

@misc{eykholt2018robustphysicalworldattacksdeep,
      title={Robust Physical-World Attacks on Deep Learning Models}, 
      author={Kevin Eykholt and Ivan Evtimov and Earlence Fernandes and Bo Li and Amir Rahmati and Chaowei Xiao and Atul Prakash and Tadayoshi Kohno and Dawn Song},
      year={2018},
      eprint={1707.08945},
      archivePrefix={arXiv},
      primaryClass={cs.CR},
      url={https://arxiv.org/abs/1707.08945}, 
}

@INPROCEEDINGS{9797363,
  author={Sun, Haipei and Wu, Kun and Wang, Ting and Wang, Wendy Hui},
  booktitle={2022 IEEE 7th European Symposium on Security and Privacy (EuroS\&P)}, 
  title={Towards Fair and Robust Classification}, 
  year={2022},
  volume={},
  number={},
  pages={356-376},
  doi={10.1109/EuroSP53844.2022.00030}}

@misc{nagireddy2023functioncompositiontrustworthymachine,
      title={Function Composition in Trustworthy Machine Learning: Implementation Choices, Insights, and Questions}, 
      author={Manish Nagireddy and Moninder Singh and Samuel C. Hoffman and Evaline Ju and Karthikeyan Natesan Ramamurthy and Kush R. Varshney},
      year={2023},
      eprint={2302.09190},
      archivePrefix={arXiv},
      primaryClass={cs.LG},
      url={https://arxiv.org/abs/2302.09190}, 
}

@inproceedings{Olszewskireproducibility2023,
author = {Olszewski, Daniel and Lu, Allison and Stillman, Carson and Warren, Kevin and Kitroser, Cole and Pascual, Alejandro and Ukirde, Divyajyoti and Butler, Kevin and Traynor, Patrick},
title = {"Get in Researchers; We're Measuring Reproducibility": A Reproducibility Study of Machine Learning Papers in Tier 1 Security Conferences},
year = {2023},
isbn = {9798400700507},
publisher = {Association for Computing Machinery},
address = {New York, NY, USA},
url = {https://doi.org/10.1145/3576915.3623130},
doi = {10.1145/3576915.3623130},
pages = {3433–3459},
numpages = {27},
location = {Copenhagen, Denmark},
series = {CCS '23}
}

@article{9802763,
  author={Strobel, Martin and Shokri, Reza},
  journal={IEEE Security \& Privacy}, 
  title={Data Privacy and Trustworthy Machine Learning}, 
  year={2022},
  volume={20},
  number={5},
  pages={44-49},
  doi={10.1109/MSEC.2022.3178187}}

@article{DBLP:journals/corr/abs-2201-02265,
  author       = {Jamie Hayes and
                  Borja Balle and
                  M. Pawan Kumar},
  title        = {Learning to be adversarially robust and differentially private},
  journal      = {CoRR},
  volume       = {abs/2201.02265},
  year         = {2022},
  url          = {https://arxiv.org/abs/2201.02265},
  eprinttype    = {arXiv},
  eprint       = {2201.02265},
  bibsource    = {dblp computer science bibliography, https://dblp.org}
}

@article{mehnaz2020black,
  title={Black-box model inversion attribute inference attacks on classification models},
  author={Mehnaz, Shagufta and Li, Ninghui and Bertino, Elisa},
  journal={arXiv preprint arXiv:2012.03404},
  year={2020}
}

@inproceedings{dolan2018fine,
  title={Fine-Pruning: Defending Against Backdooring Attacks on Deep},
  author={Dolan-Gavitt, Kang Liu Brendan and Garg, Siddharth},
  booktitle={Research in Attacks, Intrusions, and Defenses: 21st International Symposium, RAID 2018, Heraklion, Crete, Greece, September 10-12, 2018, Proceedings},
  volume={11050},
  pages={273},
  year={2018},
  organization={Springer}
}

@article{lundberg2017unified,
  title={A unified approach to interpreting model predictions},
  author={Lundberg, Scott M and Lee, Su-In},
  journal={Advances in neural information processing systems},
  volume={30},
  year={2017}
}

@inproceedings{ribeiro2016should,
  title={" Why should i trust you?" Explaining the predictions of any classifier},
  author={Ribeiro, Marco Tulio and Singh, Sameer and Guestrin, Carlos},
  booktitle={Proceedings of the 22nd ACM SIGKDD international conference on knowledge discovery and data mining},
  pages={1135--1144},
  year={2016}
}

@article{lipton2018mythos,
  title={The mythos of model interpretability: In machine learning, the concept of interpretability is both important and slippery.},
  author={Lipton, Zachary C},
  journal={Queue},
  volume={16},
  number={3},
  pages={31--57},
  year={2018},
  publisher={ACM New York, NY, USA}
}

@book{barocas2023fairness,
  title={Fairness and machine learning: Limitations and opportunities},
  author={Barocas, Solon and Hardt, Moritz and Narayanan, Arvind},
  year={2023},
  publisher={MIT press}
}

@incollection{angwin2022machine,
  title={Machine bias},
  author={Angwin, Julia and Larson, Jeff and Mattu, Surya and Kirchner, Lauren},
  booktitle={Ethics of data and analytics},
  pages={254--264},
  year={2022},
  publisher={Auerbach Publications}
}

@inproceedings{mironov2017renyi,
  title={R{\'e}nyi differential privacy},
  author={Mironov, Ilya},
  booktitle={2017 IEEE 30th computer security foundations symposium (CSF)},
  pages={263--275},
  year={2017},
  organization={IEEE}
}

@inproceedings{carlini2019secret,
  title={The secret sharer: Evaluating and testing unintended memorization in neural networks},
  author={Carlini, Nicholas and Liu, Chang and Erlingsson, {\'U}lfar and Kos, Jernej and Song, Dawn},
  booktitle={28th USENIX security symposium (USENIX security 19)},
  pages={267--284},
  year={2019}
}

@inproceedings{zhang2018protecting,
  title={Protecting intellectual property of deep neural networks with watermarking},
  author={Zhang, Jialong and Gu, Zhongshu and Jang, Jiyong and Wu, Hui and Stoecklin, Marc Ph and Huang, Heqing and Molloy, Ian},
  booktitle={Proceedings of the 2018 on Asia conference on computer and communications security},
  pages={159--172},
  year={2018}
}

@inproceedings{tramer2016stealing,
  title={Stealing machine learning models via prediction {APIs}},
  author={Tram{\`e}r, Florian and Zhang, Fan and Juels, Ari and Reiter, Michael K and Ristenpart, Thomas},
  booktitle={25th USENIX security symposium (USENIX Security 16)},
  pages={601--618},
  year={2016}
}

@inproceedings{davis2006relationship,
  title={The relationship between Precision-Recall and ROC curves},
  author={Davis, Jesse and Goadrich, Mark},
  booktitle={Proceedings of the 23rd international conference on Machine learning},
  pages={233--240},
  year={2006}
}

@article{steinhardt2017certified,
  title={Certified defenses for data poisoning attacks},
  author={Steinhardt, Jacob and Koh, Pang Wei W and Liang, Percy S},
  journal={Advances in neural information processing systems},
  volume={30},
  year={2017}
}

@article{gu2017badnets,
  title={Badnets: Identifying vulnerabilities in the machine learning model supply chain},
  author={Gu, Tianyu and Dolan-Gavitt, Brendan and Garg, Siddharth},
  journal={arXiv preprint arXiv:1708.06733},
  year={2017}
}

@article{biggio2012poisoning,
  title={Poisoning attacks against support vector machines},
  author={Biggio, Battista and Nelson, Blaine and Laskov, Pavel},
  journal={arXiv preprint arXiv:1206.6389},
  year={2012}
}

@article{goodfellow2014explaining,
  title={Explaining and harnessing adversarial examples},
  author={Goodfellow, Ian J and Shlens, Jonathon and Szegedy, Christian},
  journal={arXiv preprint arXiv:1412.6572},
  year={2014}
}

@article{szegedy2013intriguing,
  title={Intriguing properties of neural networks},
  author={Szegedy, Christian and Zaremba, Wojciech and Sutskever, Ilya and Bruna, Joan and Erhan, Dumitru and Goodfellow, Ian and Fergus, Rob},
  journal={arXiv preprint arXiv:1312.6199},
  year={2013}
}

@article{hassanpour2024impact,
  title={The Impact of Generalization Techniques on the Interplay Among Privacy, Utility, and Fairness in Image Classification},
  author={Hassanpour, Ahmad and Zarei, Amir and Mallat, Khawla and de Oliveira, Anderson Santana and Yang, Bian},
  journal={arXiv preprint arXiv:2412.11951},
  year={2024}
}

@article{Peng2022FingerprintingDN,
  title={Fingerprinting Deep Neural Networks Globally via Universal Adversarial Perturbations},
  author={Zirui Peng and Shaofeng Li and Guoxing Chen and Cheng Zhang and Haojin Zhu and Minhui Xue},
  journal={2022 IEEE/CVF Conference on Computer Vision and Pattern Recognition (CVPR)},
  year={2022},
  pages={13420-13429},
  url={https://api.semanticscholar.org/CorpusID:246904661}
}

@article{yao2022improving,
  title={Improving fairness in image classification via sketching},
  author={Yao, Ruichen and Cui, Ziteng and Li, Xiaoxiao and Gu, Lin},
  journal={arXiv preprint arXiv:2211.00168},
  year={2022}
}

@inproceedings{sablayrolles2020radioactive,
  title={Radioactive data: tracing through training},
  author={Sablayrolles, Alexandre and Douze, Matthijs and Schmid, Cordelia and J{\'e}gou, Herv{\'e}},
  booktitle={International Conference on Machine Learning},
  pages={8326--8335},
  year={2020},
  organization={PMLR}
}

@article{xu2017feature,
  title={Feature squeezing: Detecting adversarial examples in deep neural networks},
  author={Xu, Weilin and Evans, David and Qi, Yanjun},
  journal={arXiv preprint arXiv:1704.01155},
  year={2017}
}

@article{holohan2019diffprivlib,
  title={Diffprivlib: the IBM differential privacy library},
  author={Holohan, Naoise and Braghin, Stefano and Mac Aonghusa, P{\'o}l and Levacher, Killian},
  journal={arXiv preprint arXiv:1907.02444},
  year={2019}
}

@inproceedings{wang2023analysis,
  title={An analysis of untargeted poisoning attack and defense methods for federated online learning to rank systems},
  author={Wang, Shuyi and Zuccon, Guido},
  booktitle={Proceedings of the 2023 ACM SIGIR International Conference on Theory of Information Retrieval},
  pages={215--224},
  year={2023}
}

@inproceedings{knauer2024phantom,
  title={Phantom: Untargeted Poisoning Attacks on Semi-Supervised Learning},
  author={Knauer, Jonathan and Rieger, Phillip and Fereidooni, Hossein and Sadeghi, Ahmad-Reza},
  booktitle={Proceedings of the 2024 on ACM SIGSAC Conference on Computer and Communications Security},
  pages={615--629},
  year={2024}
}

@article{10.1145/3595292,
author = {Oliynyk, Daryna and Mayer, Rudolf and Rauber, Andreas},
title = {I Know What You Trained Last Summer: A Survey on Stealing Machine Learning Models and Defences},
year = {2023},
issue_date = {December 2023},
publisher = {Association for Computing Machinery},
address = {New York, NY, USA},
volume = {55},
number = {14s},
issn = {0360-0300},
url = {https://doi.org/10.1145/3595292},
doi = {10.1145/3595292},
journal = {ACM Comput. Surv.},
month = jul,
articleno = {324},
numpages = {41}
}

@article{Lukas2019DeepNN,
  title={Deep Neural Network Fingerprinting by Conferrable Adversarial Examples},
  author={Nils Lukas and Yuxuan Zhang and Florian Kerschbaum},
  journal={ArXiv},
  year={2019},
  volume={abs/1912.00888},
  url={https://api.semanticscholar.org/CorpusID:208527270}
}

@article{Meng2017MagNetAT,
  title={MagNet: A Two-Pronged Defense against Adversarial Examples},
  author={Dongyu Meng and Hao Chen},
  journal={Proceedings of the 2017 ACM SIGSAC Conference on Computer and Communications Security},
  year={2017},
  url={https://api.semanticscholar.org/CorpusID:3583538}
}

@article{Chen2018BlackMarksBM,
  title={BlackMarks: Blackbox Multibit Watermarking for Deep Neural Networks},
  author={Huili Chen and Bita Darvish Rouhani and Farinaz Koushanfar},
  journal={ArXiv},
  year={2018},
  volume={abs/1904.00344},
  url={https://api.semanticscholar.org/CorpusID:90260955}
}

@article{Papernot2015DistillationAA,
  title={Distillation as a Defense to Adversarial Perturbations Against Deep Neural Networks},
  author={Nicolas Papernot and Patrick Mcdaniel and Xi Wu and Somesh Jha and Ananthram Swami},
  journal={2016 IEEE Symposium on Security and Privacy (SP)},
  year={2015},
  pages={582-597},
  url={https://api.semanticscholar.org/CorpusID:2672720}
}

@inproceedings{Kim2017InterpretabilityBF,
  title={Interpretability Beyond Feature Attribution: Quantitative Testing with Concept Activation Vectors (TCAV)},
  author={Been Kim and Martin Wattenberg and Justin Gilmer and Carrie J. Cai and James Wexler and Fernanda B. Vi{\'e}gas and Rory Sayres},
  booktitle={International Conference on Machine Learning},
  year={2017},
  url={https://api.semanticscholar.org/CorpusID:51737170}
}

@article{Hardt2016EqualityOO,
  title={Equality of Opportunity in Supervised Learning},
  author={Moritz Hardt and Eric Price and Nathan Srebro},
  journal={ArXiv},
  year={2016},
  volume={abs/1610.02413},
  url={https://api.semanticscholar.org/CorpusID:7567061}
}

@article{Maini2021DatasetIO,
  title={Dataset Inference: Ownership Resolution in Machine Learning},
  author={Pratyush Maini},
  journal={ArXiv},
  year={2021},
  volume={abs/2104.10706},
  url={https://api.semanticscholar.org/CorpusID:231609191}
}

@inproceedings{szyller2021dawn,
  title={Dawn: Dynamic adversarial watermarking of neural networks},
  author={Szyller, Sebastian and Atli, Buse Gul and Marchal, Samuel and Asokan, N},
  booktitle={Proceedings of the 29th ACM international conference on multimedia},
  pages={4417--4425},
  year={2021}
}

@inproceedings{cohen2024simple,
  title={Simple post-training robustness using test time augmentations and random forest},
  author={Cohen, Gilad and Giryes, Raja},
  booktitle={Proceedings of the IEEE/CVF Winter Conference on Applications of Computer Vision},
  pages={3996--4006},
  year={2024}
}

@article{Zhang2019TheoreticallyPT,
  title={Theoretically Principled Trade-off between Robustness and Accuracy},
  author={Hongyang Zhang and Yaodong Yu and Jiantao Jiao and Eric P. Xing and Laurent El Ghaoui and Michael I. Jordan},
  journal={ArXiv},
  year={2019},
  volume={abs/1901.08573},
  url={https://api.semanticscholar.org/CorpusID:59222747}
}

@article{Uchida2017EmbeddingWI,
  title={Embedding Watermarks into Deep Neural Networks},
  author={Yusuke Uchida and Yuki Nagai and Shigeyuki Sakazawa and Shin’ichi Satoh},
  journal={Proceedings of the 2017 ACM on International Conference on Multimedia Retrieval},
  year={2017},
  url={https://api.semanticscholar.org/CorpusID:13060737}
}

@article{Cohen2019CertifiedAR,
  title={Certified Adversarial Robustness via Randomized Smoothing},
  author={Jeremy M. Cohen and Elan Rosenfeld and J. Zico Kolter},
  journal={ArXiv},
  year={2019},
  volume={abs/1902.02918},
  url={https://api.semanticscholar.org/CorpusID:59842968}
}

@article{Abadi2016DeepLW,
  title={Deep Learning with Differential Privacy},
  author={Mart{\'i}n Abadi and Andy Chu and Ian J. Goodfellow and H. B. McMahan and Ilya Mironov and Kunal Talwar and Li Zhang},
  journal={Proceedings of the 2016 ACM SIGSAC Conference on Computer and Communications Security},
  year={2016},
  url={https://api.semanticscholar.org/CorpusID:207241585}
}

@article{Rouhani2019DeepSignsAE,
  title={DeepSigns: An End-to-End Watermarking Framework for Ownership Protection of Deep Neural Networks},
  author={Bita Darvish Rouhani and Huili Chen and Farinaz Koushanfar},
  journal={Proceedings of the Twenty-Fourth International Conference on Architectural Support for Programming Languages and Operating Systems},
  year={2019},
  url={https://api.semanticscholar.org/CorpusID:102347976}
}

@article{Zafar2015FairnessCM,
  title={Fairness Constraints: Mechanisms for Fair Classification},
  author={Muhammad Bilal Zafar and Isabel Valera and Manuel Gomez-Rodriguez and Krishna P. Gummadi},
  journal={ArXiv},
  year={2015},
  volume={abs/1507.05259},
  url={https://api.semanticscholar.org/CorpusID:8529258}
}

@article{Chen2018DeepMarksAD,
  title={DeepMarks: A Digital Fingerprinting Framework for Deep Neural Networks},
  author={Huili Chen and Bita Darvish Rouhani and Farinaz Koushanfar},
  journal={IACR Cryptol. ePrint Arch.},
  year={2018},
  volume={2018},
  pages={322},
  url={https://api.semanticscholar.org/CorpusID:4759464}
}

@inproceedings{Han2018CoteachingRT,
  title={Co-teaching: Robust training of deep neural networks with extremely noisy labels},
  author={Bo Han and Quanming Yao and Xingrui Yu and Gang Niu and Miao Xu and Weihua Hu and Ivor Wai-Hung Tsang and Masashi Sugiyama},
  booktitle={Neural Information Processing Systems},
  year={2018},
  url={https://api.semanticscholar.org/CorpusID:52065462}
}

@article{Madry2017TowardsDL,
  title={Towards Deep Learning Models Resistant to Adversarial Attacks},
  author={Aleksander Madry and Aleksandar Makelov and Ludwig Schmidt and Dimitris Tsipras and Adrian Vladu},
  journal={ArXiv},
  year={2017},
  volume={abs/1706.06083},
  url={https://api.semanticscholar.org/CorpusID:3488815}
}

@article{Ciss2017ParsevalNI,
  title={Parseval Networks: Improving Robustness to Adversarial Examples},
  author={Moustapha Ciss{\'e} and Piotr Bojanowski and Edouard Grave and Yann Dauphin and Nicolas Usunier},
  journal={ArXiv},
  year={2017},
  volume={abs/1704.08847},
  url={https://api.semanticscholar.org/CorpusID:26714567}
}

@inproceedings{Zemel2013LearningFR,
  title={Learning Fair Representations},
  author={Richard S. Zemel and Ledell Yu Wu and Kevin Swersky and Toniann Pitassi and Cynthia Dwork},
  booktitle={International Conference on Machine Learning},
  year={2013},
  url={https://api.semanticscholar.org/CorpusID:490669}
}

@article{Zhao2018XGBODIS,
  title={XGBOD: Improving Supervised Outlier Detection with Unsupervised Representation Learning},
  author={Yue Zhao and Maciej K. Hryniewicki},
  journal={2018 International Joint Conference on Neural Networks (IJCNN)},
  year={2018},
  pages={1-8},
  url={https://api.semanticscholar.org/CorpusID:52988666}
}

@article{Guo2018CounteringAI,
  title={Countering Adversarial Images using Input Transformations},
  author={Chuan Guo and Mayank Rana and Moustapha Ciss{\'e} and Laurens van der Maaten},
  journal={ArXiv},
  year={2018},
  volume={abs/1711.00117},
  url={https://api.semanticscholar.org/CorpusID:12308095}
}

@article{Cao2019IPGuardPT,
  title={IPGuard: Protecting the Intellectual Property of Deep Neural Networks via Fingerprinting the Classification Boundary},
  author={Xiaoyu Cao and Jinyuan Jia and Neil Zhenqiang Gong},
  journal={ArXiv},
  year={2019},
  volume={abs/1910.12903},
  url={https://api.semanticscholar.org/CorpusID:204960658}
}

@inproceedings{Calmon2017OptimizedPF,
  title={Optimized Pre-Processing for Discrimination Prevention},
  author={Fl{\'a}vio du Pin Calmon and Dennis Wei and Bhanukiran Vinzamuri and Karthikeyan Natesan Ramamurthy and Kush R. Varshney},
  booktitle={Neural Information Processing Systems},
  year={2017},
  url={https://api.semanticscholar.org/CorpusID:3801798}
}

@article{Canonne2020TheDG,
  title={The Discrete Gaussian for Differential Privacy},
  author={Cl{\'e}ment L. Canonne and Gautam Kamath and Thomas Steinke},
  journal={ArXiv},
  year={2020},
  volume={abs/2004.00010},
  url={https://api.semanticscholar.org/CorpusID:214743526}
}

@article{Lai2019RobustSR,
  title={Robust Subspace Recovery Layer for Unsupervised Anomaly Detection},
  author={Chieh-Hsin Lai and Dongmian Zou and Gilad Lerman},
  journal={ArXiv},
  year={2019},
  volume={abs/1904.00152},
  url={https://api.semanticscholar.org/CorpusID:90262267}
}

@article{jiang2020robust,
  title={Robust pre-training by adversarial contrastive learning},
  author={Jiang, Ziyu and Chen, Tianlong and Chen, Ting and Wang, Zhangyang},
  journal={Advances in neural information processing systems},
  volume={33},
  pages={16199--16210},
  year={2020}
}

@article{Ross2017RightFT,
  title={Right for the Right Reasons: Training Differentiable Models by Constraining their Explanations},
  author={Andrew Slavin Ross and Michael C. Hughes and Finale Doshi-Velez},
  journal={ArXiv},
  year={2017},
  volume={abs/1703.03717},
  url={https://api.semanticscholar.org/CorpusID:7053611}
}

@article{Lukas2021SoKHR,
  title={SoK: How Robust is Image Classification Deep Neural Network Watermarking?},
  author={Nils Lukas and Edward Jiang and Xinda Li and Florian Kerschbaum},
  journal={2022 IEEE Symposium on Security and Privacy (SP)},
  year={2021},
  pages={787-804},
  url={https://api.semanticscholar.org/CorpusID:236975869}
}

@inproceedings{Adi2018TurningYW,
  title={Turning Your Weakness Into a Strength: Watermarking Deep Neural Networks by Backdooring},
  author={Yossi Adi and Carsten Baum and Moustapha Ciss{\'e} and Benny Pinkas and Joseph Keshet},
  booktitle={USENIX Security Symposium},
  year={2018},
  url={https://api.semanticscholar.org/CorpusID:3322503}
}

@inproceedings{yang2022not,
  title={Not all poisons are created equal: Robust training against data poisoning},
  author={Yang, Yu and Liu, Tian Yu and Mirzasoleiman, Baharan},
  booktitle={International Conference on Machine Learning},
  pages={25154--25165},
  year={2022},
  organization={PMLR}
}

@article{Li2021NeuralAD,
  title={Neural Attention Distillation: Erasing Backdoor Triggers from Deep Neural Networks},
  author={Yige Li and Nodens Koren and L. Lyu and Xixiang Lyu and Bo Li and Xingjun Ma},
  journal={ArXiv},
  year={2021},
  volume={abs/2101.05930},
  url={https://api.semanticscholar.org/CorpusID:231627799}
}

@article{Hong2020OnTE,
  title={On the Effectiveness of Mitigating Data Poisoning Attacks with Gradient Shaping},
  author={Sanghyun Hong and Varun Chandrasekaran and Yigitcan Kaya and Tudor Dumitras and Nicolas Papernot},
  journal={ArXiv},
  year={2020},
  volume={abs/2002.11497},
  url={https://api.semanticscholar.org/CorpusID:211506328}
}

@article{Wu2021AdversarialNP,
  title={Adversarial Neuron Pruning Purifies Backdoored Deep Models},
  author={Dongxian Wu and Yisen Wang},
  journal={ArXiv},
  year={2021},
  volume={abs/2110.14430},
  url={https://api.semanticscholar.org/CorpusID:239998081}
}

@article{Yang2020RandomizedSO,
  title={Randomized Smoothing of All Shapes and Sizes},
  author={Greg Yang and Tony Duan and J. Edward Hu and Hadi Salman and Ilya P. Razenshteyn and Jungshian Li},
  journal={ArXiv},
  year={2020},
  volume={abs/2002.08118},
  url={https://api.semanticscholar.org/CorpusID:211171876}
}

@article{Zheng2019EfficientAT,
  title={Efficient Adversarial Training With Transferable Adversarial Examples},
  author={Haizhong Zheng and Ziqi Zhang and Juncheng Gu and Honglak Lee and Atul Prakash},
  journal={2020 IEEE/CVF Conference on Computer Vision and Pattern Recognition (CVPR)},
  year={2019},
  pages={1178-1187},
  url={https://api.semanticscholar.org/CorpusID:209501025}
}

@inproceedings{Shafahi2019AdversarialTF,
  title={Adversarial Training for Free!},
  author={Ali Shafahi and Mahyar Najibi and Amin Ghiasi and Zheng Xu and John P. Dickerson and Christoph Studer and Larry S. Davis and Gavin Taylor and Tom Goldstein},
  booktitle={Neural Information Processing Systems},
  year={2019},
  url={https://api.semanticscholar.org/CorpusID:139102395}
}

@article{duddu2024combining,
  title={Combining Machine Learning Defenses without Conflicts},
  author={Duddu, Vasisht and Zhang, Rui and Asokan, N},
  journal={arXiv preprint arXiv:2411.09776},
  year={2024}
}

@article{Subramanian2023HaveTC,
  title={Have the cake and eat it too: Differential Privacy enables privacy and precise analytics},
  author={Rishabh Subramanian},
  journal={Journal of Big Data},
  year={2023},
  volume={10},
  pages={1-14},
  url={https://api.semanticscholar.org/CorpusID:259848534}
}

@article{Ye2022OnePD,
  title={One Parameter Defense—Defending Against Data Inference Attacks via Differential Privacy},
  author={Dayong Ye and Sheng Shen and Tianqing Zhu and B. Liu and Wanlei Zhou},
  journal={IEEE Transactions on Information Forensics and Security},
  year={2022},
  volume={17},
  pages={1466-1480},
  url={https://api.semanticscholar.org/CorpusID:247447226}
}

@article{AlvarezMelis2018TowardsRI,
  title={Towards Robust Interpretability with Self-Explaining Neural Networks},
  author={David Alvarez-Melis and T. Jaakkola},
  journal={ArXiv},
  year={2018},
  volume={abs/1806.07538},
  url={https://api.semanticscholar.org/CorpusID:49324194}
}

@article{Lin2023DifferentiallyPS,
  title={Differentially Private Synthetic Data via Foundation Model APIs 1: Images},
  author={Zi-Han Lin and Sivakanth Gopi and Janardhan Kulkarni and Harsha Nori and Sergey Yekhanin},
  journal={ArXiv},
  year={2023},
  volume={abs/2305.15560},
  url={https://api.semanticscholar.org/CorpusID:258888127}
}

@inproceedings{Li2023ReconstructiveNP,
  title={Reconstructive Neuron Pruning for Backdoor Defense},
  author={Yige Li and Xixiang Lyu and Xingjun Ma and Nodens Koren and L. Lyu and Bo Li and Yugang Jiang},
  booktitle={International Conference on Machine Learning},
  year={2023},
  url={https://api.semanticscholar.org/CorpusID:258865980}
}

@article{Cohen2021SimplePR,
  title={Simple Post-Training Robustness using Test Time Augmentations and Random Forest},
  author={Gilad Cohen and Raja Giryes},
  journal={2024 IEEE/CVF Winter Conference on Applications of Computer Vision (WACV)},
  year={2021},
  pages={3984-3994},
  url={https://api.semanticscholar.org/CorpusID:244709418}
}

@inproceedings{tran2024effects,
  title={On the effects of fairness to adversarial vulnerability},
  author={Tran, Cuong and Zhu, Keyu and Van Hentenryck, Pascal and Fioretto, Ferdinando},
  booktitle={Proceedings of the Thirty-Third International Joint Conference on Artificial Intelligence},
  pages={521--529},
  year={2024}
}

@INPROCEEDINGS{duddu2024sok,
  author={Duddu, Vasisht and Szyller, Sebastian and Asokan, N.},
  booktitle={2024 IEEE Symposium on Security and Privacy (SP)}, 
  title={SoK: Unintended Interactions among Machine Learning Defenses and Risks}, 
  year={2024},
  volume={},
  number={},
  pages={2996-3014},
  doi={10.1109/SP54263.2024.00243}}

@inproceedings{szyller2023conflicting,
  title={Conflicting interactions among protection mechanisms for machine learning models},
  author={Szyller, Sebastian and Asokan, N},
  booktitle={Proceedings of the AAAI Conference on Artificial Intelligence},
  volume={37},
  number={12},
  pages={15179--15187},
  year={2023}
}

@INPROCEEDINGS{noppel2024sok,
  author={Noppel, Maximilian and Wressnegger, Christian},
  booktitle={2024 IEEE Symposium on Security and Privacy (SP)}, 
  title={SoK: Explainable Machine Learning in Adversarial Environments}, 
  year={2024},
  volume={},
  number={},
  pages={2441-2459},
  doi={10.1109/SP54263.2024.00021}}

@article{chen2023privacy,
author = {Chen, Huiqiang and Zhu, Tianqing and Zhang, Tao and Zhou, Wanlei and Yu, Philip S.},
title = {Privacy and Fairness in Federated Learning: On the Perspective of Tradeoff},
year = {2023},
issue_date = {February 2024},
publisher = {Association for Computing Machinery},
address = {New York, NY, USA},
volume = {56},
number = {2},
issn = {0360-0300},
url = {https://doi.org/10.1145/3606017},
doi = {10.1145/3606017},
journal = {ACM Comput. Surv.},
month = sep,
articleno = {39},
numpages = {37}
}

@ARTICLE{gittens2022adversarial,
  author={Gittens, Alex and Yener, Bülent and Yung, Moti},
  journal={IEEE Access}, 
  title={An Adversarial Perspective on Accuracy, Robustness, Fairness, and Privacy: Multilateral-Tradeoffs in Trustworthy ML}, 
  year={2022},
  volume={10},
  number={},
  pages={120850-120865},
  doi={10.1109/ACCESS.2022.3218715}}

@article{ueda2024fairness,
  title={Fairness of artificial intelligence in healthcare: review and recommendations},
  author={Ueda, Daiju and Kakinuma, Taichi and Fujita, Shohei and Kamagata, Koji and Fushimi, Yasutaka and Ito, Rintaro and Matsui, Yusuke and Nozaki, Taiki and Nakaura, Takeshi and Fujima, Noriyuki and others},
  journal={Japanese Journal of Radiology},
  volume={42},
  number={1},
  pages={3--15},
  year={2024},
  publisher={Springer}
}

@article{rezaei2025fairness,
author = {Rezaei Nasab, Ali and Dashti, Maedeh and Shahin, Mojtaba and Zahedi, Mansooreh and Khalajzadeh, Hourieh and Arora, Chetan and Liang, Peng},
title = {Fairness Concerns in App Reviews: A Study on AI-Based Mobile Apps},
year = {2025},
issue_date = {February 2025},
publisher = {Association for Computing Machinery},
address = {New York, NY, USA},
volume = {34},
number = {2},
issn = {1049-331X},
url = {https://doi.org/10.1145/3690633},
doi = {10.1145/3690633},
journal = {ACM Trans. Softw. Eng. Methodol.},
month = jan,
articleno = {51},
numpages = {30}
}

@inproceedings{lewicki2023out,
author = {Lewicki, Kornel and Lee, Michelle Seng Ah and Cobbe, Jennifer and Singh, Jatinder},
title = {Out of Context: Investigating the Bias and Fairness Concerns of “Artificial Intelligence as a Service”},
year = {2023},
isbn = {9781450394215},
publisher = {Association for Computing Machinery},
address = {New York, NY, USA},
url = {https://doi.org/10.1145/3544548.3581463},
doi = {10.1145/3544548.3581463},
booktitle = {Proceedings of the 2023 CHI Conference on Human Factors in Computing Systems},
articleno = {135},
numpages = {17},
location = {Hamburg, Germany},
series = {CHI '23}
}

@article{melis2019exploiting,
  title={Exploiting Unintended Feature Leakage in Collaborative Learning},
  author={Luca Melis and Congzheng Song and Emiliano De Cristofaro and Vitaly Shmatikov},
  journal={2019 IEEE Symposium on Security and Privacy (SP)},
  year={2018},
  pages={691-706},
  url={https://api.semanticscholar.org/CorpusID:53099247}
}

@article{fredrikson2015model,
  title={Model Inversion Attacks that Exploit Confidence Information and Basic Countermeasures},
  author={Matt Fredrikson and Somesh Jha and Thomas Ristenpart},
  journal={Proceedings of the 22nd ACM SIGSAC Conference on Computer and Communications Security},
  year={2015},
  url={https://api.semanticscholar.org/CorpusID:207229839}
}

@article{yeom2018privacy,
  title={Privacy Risk in Machine Learning: Analyzing the Connection to Overfitting},
  author={Samuel Yeom and Irene Giacomelli and Matt Fredrikson and Somesh Jha},
  journal={2018 IEEE 31st Computer Security Foundations Symposium (CSF)},
  year={2017},
  pages={268-282},
  url={https://api.semanticscholar.org/CorpusID:2656445}
}

@inproceedings{choquette2021label,
  title={Label-only membership inference attacks},
  author={Choquette-Choo, Christopher A and Tramer, Florian and Carlini, Nicholas and Papernot, Nicolas},
  booktitle={International conference on machine learning},
  pages={1964--1974},
  year={2021},
  organization={PMLR}
}

@article{liu2022membership,
  title={Membership inference attacks against machine learning models via prediction sensitivity},
  author={Liu, Lan and Wang, Yi and Liu, Gaoyang and Peng, Kai and Wang, Chen},
  journal={IEEE Transactions on Dependable and Secure Computing},
  volume={20},
  number={3},
  pages={2341--2347},
  year={2022},
  publisher={IEEE}
}

@article{shokri2017membership,
  title={Membership Inference Attacks Against Machine Learning Models},
  author={R. Shokri and Marco Stronati and Congzheng Song and Vitaly Shmatikov},
  journal={2017 IEEE Symposium on Security and Privacy (SP)},
  year={2016},
  pages={3-18},
  url={https://api.semanticscholar.org/CorpusID:10488675}
}

@INPROCEEDINGS{newaz2020adversarial,
  author={Newaz, Akm Iqtidar and Haque, Nur Imtiazul and Sikder, Amit Kumar and Rahman, Mohammad Ashiqur and Uluagac, A. Selcuk},
  booktitle={GLOBECOM 2020 - 2020 IEEE Global Communications Conference}, 
  title={Adversarial Attacks to Machine Learning-Based Smart Healthcare Systems}, 
  year={2020},
  volume={},
  number={},
  pages={1-6},
  doi={10.1109/GLOBECOM42002.2020.9322472}}

@article{finlayson2019adversarial,
author = {Samuel G. Finlayson  and John D. Bowers  and Joichi Ito  and Jonathan L. Zittrain  and Andrew L. Beam  and Isaac S. Kohane },
title = {Adversarial attacks on medical machine learning},
journal = {Science},
volume = {363},
number = {6433},
pages = {1287-1289},
year = {2019},
doi = {10.1126/science.aaw4399},
URL = {https://www.science.org/doi/abs/10.1126/science.aaw4399},
eprint = {https://www.science.org/doi/pdf/10.1126/science.aaw4399}}

@Inbook{khayyam2020artificial,
author="Khayyam, Hamid
and Javadi, Bahman
and Jalili, Mahdi
and Jazar, Reza N.",
editor="Jazar, Reza N.
and Dai, Liming",
title="Artificial Intelligence and Internet of Things for Autonomous Vehicles",
bookTitle="Nonlinear Approaches in Engineering Applications: Automotive Applications of Engineering Problems",
year="2020",
publisher="Springer International Publishing",
address="Cham",
pages="39--68",
}

@article{amer2020dynamic,
title = {A dynamic Windows malware detection and prediction method based on contextual understanding of API call sequence},
journal = {Computers \& Security},
volume = {92},
pages = {101760},
year = {2020},
issn = {0167-4048},
doi = {https://doi.org/10.1016/j.cose.2020.101760},
url = {https://www.sciencedirect.com/science/article/pii/S0167404820300444},
author = {Eslam Amer and Ivan Zelinka}
}

@article{apruzzese2023role,
author = {Apruzzese, Giovanni and Laskov, Pavel and Montes de Oca, Edgardo and Mallouli, Wissam and Brdalo Rapa, Luis and Grammatopoulos, Athanasios Vasileios and Di Franco, Fabio},
title = {The Role of Machine Learning in Cybersecurity},
year = {2023},
issue_date = {March 2023},
publisher = {Association for Computing Machinery},
address = {New York, NY, USA},
volume = {4},
number = {1},
url = {https://doi.org/10.1145/3545574},
doi = {10.1145/3545574},
journal = {Digital Threats},
month = mar,
articleno = {8},
numpages = {38}
}

@incollection{sirignano2021universal,
  title={Universal features of price formation in financial markets: perspectives from deep learning},
  author={Sirignano, Justin and Cont, Rama},
  booktitle={Machine learning and AI in finance},
  pages={5--15},
  year={2021},
  publisher={Routledge}
}

@article{bravi2024development,
  title={Development and use of machine learning algorithms in vaccine target selection},
  author={Bravi, Barbara},
  journal={npj Vaccines},
  volume={9},
  number={1},
  pages={15},
  year={2024},
  publisher={Nature Publishing Group UK London}
}

@Article{rahmani2021machine,
AUTHOR = {Rahmani, Amir Masoud and Yousefpoor, Efat and Yousefpoor, Mohammad Sadegh and Mehmood, Zahid and Haider, Amir and Hosseinzadeh, Mehdi and Ali Naqvi, Rizwan},
TITLE = {Machine Learning (ML) in Medicine: Review, Applications, and Challenges},
JOURNAL = {Mathematics},
VOLUME = {9},
YEAR = {2021},
NUMBER = {22},
ARTICLE-NUMBER = {2970},
URL = {https://www.mdpi.com/2227-7390/9/22/2970},
ISSN = {2227-7390},
DOI = {10.3390/math9222970}
}
\newpage

\appendix

\definecolor{prestage}{HTML}{FFF5E6}     
\definecolor{duringstage}{HTML}{F0FAF0}  
\definecolor{poststage}{HTML}{F0F8FF}    
\definecolor{deploystage}{HTML}{F8F0FF}  
\definecolor{headercolor}{HTML}{F8F9FA}  

\begin{table*}[t]
\caption{ The table summarizes surveyed trustworthy ML techniques in \textsc{Landseer}, including the technique type, application stage, referenced work, publication year, key properties, and composability notes. Properties capture artifact availability (A) and whether it is official or third-party, reproducibility (R), replicability (P), self-contained (S), extensibility (E), modularity (M), and integration into Landseer (I). Symbols denote: \faCircle[regular] (absent), \faCircle (present or official artifact), \faAdjust (partially present or third-party artifact), and \textcolor{integrated}{\faCog} (integrated in Landseer). Properties S, E, and M are aggregated: \faCircle if all are present, \faAdjust if partially satisfied, and \faCircle[regular] if all are absent.
}

\label{tbl:litSurveyFA}
\centering
\footnotesize
\setlength{\tabcolsep}{3pt}
\renewcommand{\arraystretch}{0.8}
\begin{tabular}{>{\centering\arraybackslash}p{0.6cm}|>{\centering\arraybackslash}p{1.3cm}|>{\raggedright\arraybackslash}p{3.3cm}|>{\centering\arraybackslash}p{0.6cm}|>{\raggedright\arraybackslash}p{2.0cm}|>{\raggedright\arraybackslash}p{8.0cm}}
\toprule
\rowcolor{headercolor}
\textbf{Type} & \textbf{Stage} & \textbf{Citation} & \textbf{Year} & \textbf{Properties (A,R,P,[S,E,M],I)} & \textbf{Non-Composability Reasons / Composability Notes} \\
\midrule
\multirow{6}{*}{\atdefense} & \stagepre & Cisse et al.~\cite{Ciss2017ParsevalNI} & 2017 & \faCircle[regular]\faCircle[regular]\faCircle[regular]\faCircle[regular]  &  Not composable due to the absence of a publicly available artifact\\
\cmidrule{2-6}
    & \multirow{3}{*}{\stageduring} & Madry et al.~\cite{Madry2017TowardsDL} & 2017 & \faCircle\faCircle\faCircle\faCircle[regular] &  The framework is incompatible with the other tools selected for the experiment.\\
\cmidrule{3-6}
& & Shafahi et al.~\cite{Shafahi2019AdversarialTF} & 2019 & \faCircle\faCircle\faCircle\faCircle[regular] &  The framework is incompatible with the other tools selected for the experiment.\\
\cmidrule{3-6}
& & Zhang et al.~\cite{Zhang2019TheoreticallyPT} (\AR{in}{})& 2019 & \faCircle\faCircle[regular]\faCircle\faCircle\textcolor{integrated}{\faCog}   & Not reproducible due to CUDA 9 dependency. Replicated via official artifact.\\
\cmidrule{2-6}
& \multirow{2}{*}{\stagepost} & Papernot et al.~\cite{Papernot2015DistillationAA} & 2015 & \faCircle[regular]\faCircle[regular]\faCircle[regular]\faCircle[regular] & Not composable due to the absence of a publicly available artifact \\
\cmidrule{3-6}
& & Meng et al.~\cite{Meng2017MagNetAT} (\AR{post}{})& 2017 & \faCircle\faCircle\faCircle\faCircle\textcolor{integrated}{\faCog}  &  Containerized with minimal efforts\\
\midrule
\multirow{5}{*}{\ordefense} & \multirow{3}{*}{\stagepre} & Lai et al.~\cite{Lai2019RobustSR} & 2019 & \faCircle\faCircle\faCircle[regular]\faCircle[regular] & Reproduced on official datasets but not replicated on CIFAR-10; not composable in our setup. \\
\cmidrule{3-6}
& & Zhao et al.~\cite{Zhao2018XGBODIS} (\OR{pre}{}) & 2018 & \faCircle\faCircle\faCircle\faCircle\textcolor{integrated}{\faCog}  &  Added an adapter that runs outlier filtering and writes standardized output npy data artifacts for downstream tools. \\
\cmidrule{3-6}
& & Borgnia et al.~\cite{Borgnia2020StrongDA} & 2020 & \faCircle[regular]\faCircle[regular]\faCircle[regular]\faCircle[regular] & Not composable due to the absence of a publicly available artifact \\
\cmidrule{2-6}
& \stageduring & Han et al.~\cite{Han2018CoteachingRT} (\OR{in}{}) & 2018 & \faCircle\faCircle\faCircle\faCircle\textcolor{integrated}{\faCog} & Changed python2 to python3\\
\cmidrule{2-6}
& \stagepost & Li et al.~\cite{Li2023ReconstructiveNP}  (\OR{post}{}) & 2023 & \faCircle\faCircle\faCircle\faCircle\textcolor{integrated}{\faCog}  & Containerized with minimal effort\\ 
\midrule
\multirow{7}{*}{\wmdefense} & \stagepre & Sablayrolles et al.~\cite{sablayrolles2020radioactive} & 2020 & \faCircle\faCircle[regular]\faCircle[regular]\faCircle[regular] & Irrecoverable issues with codebase and reproducing results\\
\cmidrule{3-6}
& & Gu et al. ~\cite{gu2017badnets} (\WM{pre}{}) & 2017 & \faCircle\faCircle\faCircle\faAdjust\textcolor{integrated}{\faCog}  & Reworked the data pipeline to pre-generate backdoored datasets, decoupling poisoning from training \\
\cmidrule{2-6}
 & \multirow{4}{*}{\stageduring} & Adi et al.~\cite{Adi2018TurningYW} & 2018 & \faCircle\faCircle\faCircle[regular]\faCircle[regular] & Challenging to containerize as it has an outdated dependency stack, including specific versions of PyTorch and CUDA.  \\ 
\cmidrule{3-6}
& & Rouhani et al.~\cite{Rouhani2019DeepSignsAE} & 2019 & \faCircle[regular]\faCircle[regular]\faCircle[regular]\faCircle[regular] & Not composable due to the absence of a publicly available artifact \\
\cmidrule{3-6}
& & Uchida et al.~\cite{Uchida2017EmbeddingWI} (\WM{in}{})& 2017 & \faCircle\faCircle\faCircle\faCircle\textcolor{integrated}{\faCog} & Changed Keras to Pytorch \\
\cmidrule{3-6}
& & Zhang et al.~\cite{zhang2018protecting} & 2018 & \faCircle[regular]\faCircle[regular]\faCircle[regular]\faCircle[regular] & Not composable due to the absence of a publicly available artifact  \\
\cmidrule{2-6}
& \stagepost & Chen et al.~\cite{Chen2018BlackMarksBM} & 2018 & \faCircle[regular]\faCircle[regular]\faCircle[regular]\faCircle[regular] & Not composable due to the absence of a publicly available artifact \\
\cmidrule{2-6}
& \stagedeploy & Szyller et al.~\cite{szyller2021dawn} & 2021 & \faCircle\faCircle\faCircle\faCircle\textcolor{integrated}{\faCog} & Changed pickle (.pkl) to Numpy .npy format and containerized \\
\midrule
\multirow{4}{*}{\fpdefense} & \stageduring & Chen et al.~\cite{Chen2018DeepMarksAD} & 2018 & \faCircle[regular]\faCircle[regular]\faCircle[regular]\faCircle[regular] & Not composable due to the absence of a publicly available artifact  \\
\cmidrule{2-6}

& \multirow{3}{*}{\stagedeploy} & Lukas et al.~\cite{Lukas2019DeepNN} & 2019 & \faCircle[regular]\faCircle[regular]\faCircle[regular]\faCircle[regular] & Not composable due to the absence of a publicly available artifact  \\
\cmidrule{3-6}
& & Maini et al.~\cite{Maini2021DatasetIO}  (\FP{dep}{})& 2021 & \faCircle\faCircle\faCircle\faCircle\textcolor{integrated}{\faCog} & Containerized with minimal effort\\ 
\cmidrule{3-6}
& & Cao et al.~\cite{Cao2019IPGuardPT} & 2019 & \faCircle[regular]\faCircle[regular]\faCircle[regular]\faCircle[regular] & Not composable due to the absence of a publicly available artifact  \\

\midrule
\multirow{6}{*}{\dpdefense} & \multirow{3}{*}{\stagepre} & Lin et al.~\cite{Lin2023DifferentiallyPS} & 2023 & \faCircle\faCircle[regular]\faCircle[regular]\faCircle[regular] & Irrecoverable codebase error \\
\cmidrule{3-6}
& & Canonne et al.~\cite{Canonne2020TheDG} & 2020 & \faCircle\faCircle\faCircle[regular]\faCircle[regular] & Unable to replicate on CIFAR-10 due to limited pipeline documentation and integration guidance \\
\cmidrule{3-6}
& & Subramanian et al.~\cite{Subramanian2023HaveTC} & 2023 & \faCircle[regular]\faCircle[regular]\faCircle[regular]\faCircle[regular] & Not composable due to the absence of a publicly available artifact  \\
\cmidrule{2-6}
& \stagepost & Ye et al.~\cite{Ye2022OnePD} & 2022 & \faCircle[regular]\faCircle[regular]\faCircle[regular]\faCircle[regular] & Not composable due to the absence of a publicly available artifact \\
\cmidrule{2-6}
& \stageduring & Holohan et al.~\cite{holohan2019diffprivlib} & 2019 & \faCircle\faCircle[regular]\faCircle\faCircle[regular] & Unable to reproduce; CIFAR-10 replication <50\% (CNN) and \textasciitilde{35}\% (ResNet), thus excluded.\\
\cmidrule{2-6}
 & \stageduring, \stagedeploy & Abadi et al.~\cite{Abadi2016DeepLW}  (\DP{in}{}) (\DP{dep}{}) & 2016 & \faCircle\faCircle[regular]\faCircle\faCircle\textcolor{integrated}{\faCog}  & Unable to reproduce with ResNet18 (Opacus compatibility) but replicated with ResNet20 (no BatchNorm) \\
\midrule
\multirow{6}{*}{\fairnessdefense} & \multirow{2}{*}{\stagepre} & Calmon et al.~\cite{Calmon2017OptimizedPF} & 2017 & \faCircle\faCircle[regular]\faCircle[regular] \faCircle[regular]  & Not composable due to dataset incompatibility (tabular data) \\
\cmidrule{3-6}
& & Zemel et al.~\cite{Zemel2013LearningFR} & 2013 & \faAdjust\faCircle[regular]\faCircle[regular] \faCircle[regular] & Not composable due to dataset incompatibility (tabular data)   \\
\cmidrule{2-6}
 & \multirow{3}{*}{\stageduring} & Zafar et al.~\cite{Zafar2015FairnessCM} & 2015 & \faCircle\faCircle[regular]\faCircle[regular] \faCircle[regular]  & Not composable due to dataset incompatibility (tabular data) \\
\cmidrule{3-6}
& & Yao et al.~\cite{yao2022improving}  \GF{in}{} & 2022 & \faCircle\faCircle\faCircle\faAdjust\textcolor{integrated}{\faCog} & Modified dataset split to enable DP/DEO (original test set single-class); integrated sketch preprocessing as subprocess. \\
\cmidrule{2-6}
& \stagedeploy & Hardt et al.~\cite{Hardt2016EqualityOO} & 2016 & \faAdjust\faCircle[regular]\faCircle[regular]\faCircle[regular]  & Not composable due to dataset incompatibility (tabular data) \\
\midrule
\multirow{2}{*}{\expdefense }& \stageduring & Alvarez-Melis et al.~\cite{AlvarezMelis2018TowardsRI} & 2018 & \faCircle[regular]\faCircle[regular]\faCircle[regular]\faCircle[regular] & Not composable due to the absence of a publicly available artifact \\
\cmidrule{2-6}
 & \stagedeploy & Kim et al.~\cite{Kim2017InterpretabilityBF} & 2017 & \faCircle\faCircle\faCircle[regular]\faCircle[regular]& Unable to replicate for CIFAR-10 due to missing required files \\
 \cmidrule{3-6}
 & & Lundberg et al. ~\cite{ExplainShap_NIPS2017_8a20a862}  \EX{dep}{} & 2017 & \faCircle\faCircle\faCircle\faCircle\textcolor{integrated}{\faCog} & Applied SHAP to replicate the technique on CIFAR-10 \\
\bottomrule
\end{tabular}
\end{table*}
\begin{table}[H]
    \begin{footnotesize}
    \centering
    \caption{\footnotesize Trustworthy ML techniques and hyperparameters used}
    \label{tbl:hyperparameters}
\begin{tabular}{lcc}
    \toprule
    \textbf{Defense Technique} &\textbf{ Hyperparameter} &\textbf{ Value} \\
    \midrule
    \WM{pre}{} & Watermarking fraction & 0.1 \\
    \midrule
    \AR{in}{} & Trade-off regularization & 6 \\
    \midrule
    \AR{post}{} & Weight decay regularization & $e^{-9}$ \\
                & Autoencoder noise-level & 0.025\\
    \midrule
    \OR{pre}{} & Contamination rate & 0.1 \\
              & Number of estimators & 20 \\

    \midrule
    \OR{in}{} & Forget rate & 0.5 \\
              & Noise type & symmetric \\

    \midrule
    \OR{post}{} & Prunning threshold & 0.1 \\

    \midrule
    \WM{pre}{} & Trigger size & 5x5\\
               & Watermarking fraction & 0.1 \\

    \midrule
    \WM{in}{} & Watermark strength  & 0.01 \\

    \midrule
    \FP{dep}{} & Num. of fingerprints  & 100 \\
                & Significance level ($\alpha$) & 0.01 \\

    \midrule
    \DP{in}{} & Maximum gradient norm & 2.0 \\
            & Target epsilon (delta) & 50 ($e^{-5}$) \\

    \midrule
    \DP{dep}{} & Target epsilon (delta) & 50 ($e^{-5}$) \\

    \midrule 
    \EX{dep}{} &  Num. of SHAP samples & 500 \\
    
    \bottomrule
\end{tabular}
    \end{footnotesize}
\end{table}

\section{Literature distribution}

The Table below provides a breakdown of the venues where the work was selected.

\begin{table}[H]
    \begin{footnotesize}
    \centering
    \caption{\footnotesize Selected venues and tally of work collected}
    \label{tbl:papers}
\begin{tabular}{lcc}
    \toprule
    Conference & Acronym & No. \\
    \midrule
    Conf. on Neural Information Processing Sys. & NeurIPS  & 11 \\ 
    Int. Conf. on Machine Learning & ICML & 09 \\ 
    Int. Conf. on Learning Representations & ICLR & 08 \\ 
    Conf. on Comp. and Communications Security & CCS & 03 \\
    Int. Conf. on Multimedia Retrieval & ICMR & 03 \\
    Conf. on Comp. Vision and Pattern Recognition & CVPR & 02 \\
    Int. Joint Conference on Artificial Intelligence & IJCAI & 02 \\ 
    USENIX Security & Security & 02 \\ 
    Arch. Supp. for Prog. Lang. and Oper. Systems  & ASPLOS & 01\\
    Symp. on Security and Privacy & S\&P & 01 \\
    Netw. and Distrib. Syst. Security & NDSS & 01 \\
    Winter Conf. on Appl. of Comp. Vision & WACV & 01 \\
    Int. Joint Conf. on Neural Networks & IJCNN & 01 \\
    Int. Conf. on Acoust., Speech, and Signal Process. & ICASSP & 01 \\
    Assoc. for the Adv. of Artificial Intelligence & AAAI & 01 \\  
    J. of Big Data & JBD & 01 \\
    Trans. on Inf. Forensics and Security & TIFS & 01 \\
    Int. Conf. on Knowl. Discov. and Data Min. & KDD & 01 \\
    \midrule
    Grey literature additions & - & 04 \\
    Total & - & 54 \\
    \bottomrule
\end{tabular}
    \end{footnotesize}
\end{table}

\section{Open Science}
The tools utilized in this work are available in their respective repositories, and the datasets are well-known and easy to access through various platforms (e.g. huggingface). 
We make the Landseer code available at https://anonymous.4open.science/r/Landseer-FF4A for reviewers to peruse. 
If accepted, we aim to make the Landseer framework public under MIT License.

\section{Ethical Considerations}
This work does not substantially affect the status quo of current practices in \defenses.
In particular, the defenses and datasets are public knowledge and are widely used in the field, and we do not foresee ethical effects of their use in this work.
We expect that this work will foster the adoption and integration of ethically relevant tools to achieve fairness, privacy, and robustness.

\section {Generative AI Usage}
\noindent\textbf{System Implementation}: We used GitHub Copilot to assist with the development of the \landseer framework and replicating some \defenses. We manually verified the correctness of all the code generated by the tool. Our manual evaluation of \landseer and the respective defenses also serves to validate the correctness of the generated code.

\noindent\textbf{Writing}: We used Grammarly to help with editorial edits like grammar, spelling, and simple rephrasing. We verified all factual statements in the paper and ensured that no content generated by AI was included without verification. 
\end{document}